\documentclass[]{aa}

\usepackage{graphicx}           
\usepackage{float}               
\usepackage{placeins}             
\usepackage{adjustbox}           
\usepackage{lscape}               
\usepackage{wrapfig}             
\usepackage{epstopdf}              
\usepackage{subfigure}             
\usepackage{booktabs}              
\usepackage{amsmath}               
\usepackage[T1]{fontenc}           
\usepackage{float}                 
\usepackage{gensymb}             
\usepackage{rotating}             
\usepackage{siunitx}
\usepackage{dcolumn}
\usepackage{hyperref}
\usepackage{dblfloatfix}
\usepackage{adjustbox}
\usepackage{txfonts}
\usepackage{natbib}
\usepackage{graphicx}  
\usepackage{amsmath}   
\usepackage{multirow}  

\bibpunct{(}{)}{;}{a}{}{,}

\DeclareRobustCommand{\VAN}[3]{#2}
\let\VANthebibliography\thebibliography
\def\thebibliography{\DeclareRobustCommand{\VAN}[3]{##3}\VANthebibliography}
\newcommand{\chandra}{\textsl{Chandra}\xspace}
\newcommand{\nustar}{\textsl{NuSTAR}\xspace}
\newcommand{\xmmnewton}{\textsl{XMM-Newton}\xspace}
\newcommand{\swift}{ \textit{Swift}/BAT\xspace}

\DeclareGraphicsExtensions{.ps}
\usepackage{graphicx}

\usepackage{txfonts}

\begin{document}

\title{Spin-down of the  accreting magnetar candidate 
4U 0114+65: possible first evidence for a strong coupling regime}

\author{
Sanjurjo-Ferr\'{i}n, G.$^{1}$,
Torrej\'on, J.M.$^{1}$,
Postnov, K.$^{2}$,
Rodes-Roca, J.J.$^{1}$,
Oskinova, L.$^{3}$,
Planelles-Villalva, J.$^{1}$
}

\institute{
$^{1}$Instituto Universitario de F\'{i}sica Aplicada a las Ciencias y las Tecnolog\'ias, Universidad de Alicante, 03690 Alicante, Spain\\
$^{2}$Sternberg Astronomical Institute, Moscow M.V. Lomonosov State University, Universitetskij pr, 13, Moscow 119234, Russia\\
$^{3}$Institute for Physics and Astronomy, Universit\"{a}t Potsdam, 14476 Potsdam, Germany\\
}

\date{Received XXX / Accepted XXX}

\abstract
{4U~0114+65 is a high-mass X-ray binary composed of the B1\,Ia supergiant V{*}~V662~Cas and one of the slowest known accreting neutron stars, with a spin period of $\sim$9.4 ks. In 2025, the source entered an unusual state in which its long X-ray pulsations became undetectable in the \textit{Swift}/BAT monitoring for at least several months, motivating a Director's Discretionary Time observation with \textit{XMM-Newton}.
We compare this 2025 observation with a previous \textit{XMM-Newton} observation obtained in 2015, when the source was brighter and clearly pulsed. We also analyze the long-term evolution of the spin period using \textit{Swift}/BAT data. The spectra were studied through average and pulse-phase-resolved analyses using the same phenomenological model adopted in previous work.
We find that the 2025 \textit{XMM-Newton} observation still reveals weak pulsations, with a period of about 9.3 ks, despite their non-detection in the \textit{Swift}/BAT monitoring. The overall spectral shape remains similar in both epochs, but the source luminosity decreased by about one order of magnitude, mainly because of a strong suppression of the bulk-motion Comptonization component. Although the absorbing column is higher in 2025, the inferred wind properties remain broadly compatible with those obtained in 2015, suggesting that no major global change in the donor wind is required. Instead, the results point to a substantial reduction in the accretion efficiency close to the NS magnetosphere.
We propose that 4U~0114+65 may be evolving toward a regime of partial centrifugal inhibition in the strong-coupling regime, when the toroidal magnetic-field component is comparable to the poloidal one. If true, this phenomenon has been observed for the first time. In this state, accretion becomes progressively less efficient and more intermittent without reaching a fully developed propeller regime. In this scenario, the apparent vanishing of the pulse in long-term hard X-ray monitoring would be a consequence of reduced luminosity and lower absolute pulsed flux, rather than a complete disappearance of the underlying spin modulation.}


\titlerunning{Spin-down of the accreting magnetar 4U 0114+65}

\authorrunning{Sanjurjo-Ferr\'{\i}n et al. }

\maketitle

\section{Introduction}
High-mass X-ray binaries (HMXRBs) host a compact object (a neutron star, NS, or a black hole) accreting from a massive stellar companion. They constitute prime laboratories to investigate accretion physics and the structure of massive-star winds \citep{MartinezNunez2017}. Beyond their relevance for compact-object formation channels and the evolution of close binaries---potential progenitors of compact-object mergers and associated gravitational-wave and short $\gamma$-ray burst events---HMXRBs also probe the behavior of matter under extreme gravitational and magnetic fields.

\object{4U\,0114+65} was discovered in the SAS~3 Galactic survey \citep{1977IAUC.3144....2D}. The donor is the luminous B1\,Ia supergiant \object{V$^{*}$~V662~Cas} \citep{Reig}. With an orbital period of $\sim$11.6~d, the NS orbits deeply embedded in the stellar wind, at an orbital separation of $\approx 1.34$--$1.65\,R_{\star}$, providing a direct window onto the inner wind region of a B1 supergiant. The system is non-eclipsing \citep{2015MNRAS.454.4467P} and displays pronounced temporal and spectral variability over a wide range of timescales \citep{1985ApJ...299..839C,2017ApJ...844...16H}. A summary of the main system parameters is given in Table~\ref{4u_parameters}.

 A defining property of \object{4U\,0114+65} is its unusually slow X-ray pulsations, with $P_{\rm spin}\simeq 9.36$~ks, placing it among the slowest known accreting pulsars. The spin period of \object{4U\,0114+65} has evolved rapidly over the past decades. It was first measured at $\sim$10.01~ks \citep{1992A&A...262L..25F} and subsequently reported at $\sim$9.83~ks \citep{2000ApJ...536..450H}, $\sim$9.61~ks \citep{2005A&A...436L..31B}, and $\sim$9.54~ks \citep{2006MNRAS.367.1457F}. \citet{2011MNRAS.413.1083W} documented a further decrease from $\sim$9.61~ks to $\sim$9.47~ks between 2003 and 2008, corresponding to a spin-up rate of $\sim 1.09\times 10^{-6}\,\mathrm{s^{-1}}$, and \citet{2017A&A...606A.145S} reported $P_{\rm spin}\simeq 9.36$~ks. Such long spin periods have been discussed as potential indicators of strong magnetic fields in accreting NSs \citep{2013ARep...57..287I}. In this context, \citet{1999ApJ...513L..45L} and \citet{2017A&A...606A.145S} suggested that the NS could have been born as a magnetar. Magnetars are NSs with magnetic fields $B\simeq 10^{14}$--$10^{15}$~G \citep{1992ApJ...392L...9D}. Magnetar-like objects may exist in HMXRBs \citep{2008ApJ...683.1031B} accreting from the wind of the massive companion \citep{2012MNRAS.425..595R}, the so called accreting magnetars. As an alternative interpretation, \citet{2006A&A...458..513K} proposed that the observed pulsations could arise from accretion from a structured wind shaped by tidally driven oscillations in the supergiant photosphere induced by the close NS orbit.

On longer timescales, the source exhibits super-orbital modulations and torque reversals, pointing to complex interactions between the NS and the accreting flow \citep{2006MNRAS.367.1457F,2006AdSpR..38.2779S,2017A&A...606L..10B}. \citet{2017ApJ...844...16H} proposed that the changing spin behavior could be driven by the formation of a transient accretion disk. Within a wind-accretion framework, \citet{2017A&A...606A.145S} interpreted the spin evolution (its duration and rapid changes) in terms of quasi-spherical settling accretion \citep{2012MNRAS.420..216S}. In this regime, typical of moderately luminous systems (below $\simeq 4\times 10^{36}$~erg~s$^{-1}$), a convective quasi-spherical shell forms above the NS magnetosphere, and the plasma entry rate through the magnetosphere is regulated by the dominant cooling channel of the hot shell, primarily Compton and radiative losses.
Recent pointed observations with \nustar\ and \xmmnewton\ have further highlighted the complexity of \object{4U\,0114+65}, reporting signatures consistent with a co-rotating interaction region (CIR), a mechanism commonly invoked to explain periodic variability in UV wind lines of B-type supergiants \citep{2008ApJ...678..408L} and X-ray modulation in O stars \citep{2001A&A...378L..21O,2014MNRAS.441.2173M}, as well as off-states and potential cyclotron resonant scattering features (CRSFs) \citep{2023MNRAS.522.3271A}, which could challenge a magnetar interpretation. Moreover, \citet{2025A&A...694A.192S} argued that the soft excess around $\sim$2~keV is produced, at least in part, by unresolved line emission, becoming apparent mainly in the most absorbed spectra when stellar wind suppresses the NS soft continuum. The same study connected the pronounced pulse-to-pulse variability and short-lived flares to quasi-spherical settling accretion cycles, where matter accumulates near the magnetospheric boundary until efficient Compton cooling triggers a bright phase, followed by gradual depletion and the onset of a new cycle. Short-timescale features in the light curves have been associated with different physical phenomena. In \citet{2017A&A...606A.145S}, dips consistent with the passage of dense clumps and low-density inter-clump regions were reported. More recently, \citet{2025A&A...694A.192S} interpreted short-duration spikes ($\sim$0.1--0.5\,ks) as possible signatures of Rayleigh--Taylor instabilities (RTIs) operating at the magnetospheric boundary, while longer-duration spikes were associated with variability close to the Bondi radius.

In 2025, 4U~0114+65 entered a rare state in which its pulsations became undetectable in the \textit{Swift}/BAT monitoring and remained so for at least the subsequent six months, with an unknown recovery time. While off-states are commonly observed in accreting pulsars, the sustained disappearance of pulsations over such a long interval is highly unusual. The loss of pulsations is clearly apparent in \textit{Swift}/BAT data: a sliding-window timing analysis around $P_{\rm spin}$ confirms the long-term presence and evolution of the pulse while revealing intermittent gaps, which have grown substantially since 2022, reaching durations exceeding one year for the first time. This new state has not yet been examined with the spectral and timing capabilities of large X-ray telescopes. This motivated the Director's Discretionary Time observation that is the subject of the present work.

\begin{table}
\caption{Main astrophysical parameters of \object{4U\,0114+65}.}
\label{4u_parameters}
\centering

\begin{tabular}{ll}
\hline\hline
Parameter & Value \\
\hline
\multicolumn{2}{l}{\textit{Companion}$^{a}$} \\
\hline
Spectral type                & B1\,Ia \\
$T_{\rm eff}$ (K)            & $24000\pm3000$ \\
Radius ($R_{\odot}$)         & $37\pm15$ \\
Mass ($M_{\odot}$)           & $16\pm5$ \\
$M_{V}$                        & $-7\pm1$ \\
$E(B-V)$                        & $1.24\pm0.02$ \\
$BC$                           & $-2.3\pm0.3$ \\
$M_{\rm bol}$                  & $-9.3\pm1.0$ \\
$v\sin i$                      & $96\pm20$ \\
$v_{\infty}$ (km\,s$^{-1}$)  & 1200 \\
\hline
\multicolumn{2}{l}{\textit{System}}\\
\hline
Distance $d$ (kpc)$^{b}$           & $4.5^{+0.3}_{-0.2}$ \\
$P_{\rm orb}$ (d)$^{c}$            & $11.6\pm0.1$ \\
$P_{\rm superorb}$ (d)$^{d}$       & $30.7\pm0.1$ \\
$P_{\rm spin}$ (ks)$^{e}$           & $9.1-9.3$ \\
Inclination$^{f}$                  & $\sim 45^{\circ}$ \\
Eccentricity$^{g}$                 & $0.18\pm0.05$ \\
Argument of periastron$^{c}$       & $11\pm11^{\circ}$ \\
v$_{rel}$ (km\,s$^{-1}$)  & $\approx 560 \pm 70$ \\
$L_{X}$ ($\times 10^{36}$ erg\,s$^{-1}$)$^{e}$ & 0.1--1.3 \\
\hline
\multicolumn{2}{l}{\textit{Characteristic radii}$^{h}$}\\
\hline
$R_{\rm Bondi}$ (cm)         & $\approx 10^{11}$ \\
$R_{\rm cor}$ (cm)           & $\approx 7\times10^{10}$ \\
$R_{\rm A}$ (cm)             & $\approx 1.7\times10^{10}$ \\

\hline
\end{tabular}

\tablefoot{
$^{a}$\citet{Reig};
$^{b}$\citet{Bailer-Jones_2021};
$^{c}$\citet{grun};
$^{d}$\citet{2006MNRAS.367.1457F};
$^{e}$This work;
$^{f}$\citet{grun} and this work;
$^{g}$\citet{grun,1985ApJ...299..839C}.
$^{h}$\citet{2017A&A...606A.145S}.
The wind velocity at the NS location is estimated with a $\beta$-law,
$v_{\rm wind}=v_{\infty}\left(1-\frac{R_*}{a}\right)^{\beta}$ with $\beta=0.8$ \citep{1999isw..book.....L}.
Here $R_*$ is the donor radius and $a$ the orbital separation.
We define $R_{\rm Bondi}\approx \frac{2GM}{v_{\rm wind}^{2}+v_{\rm orb}^{2}}$,
$R_{\rm cor}\approx \left(\frac{GM P_*^{2}}{4\pi^{2}}\right)^{1/3}$,
and $R_{\rm A}$ as the magnetospheric (Alfv\'en) radius.
}
\end{table}

In this paper, we compare  two XMM-Newton observations obtained a decade apart, in 2015 and 2025 with the aim of investigating the reason for the apparent disappearance of the \swift detection of the pulse. Hereafter, we refer to the 2015 and 2025 \xmmnewton observations as XMM-1 and XMM-2, respectively.

\section{Observations and analysis}
\label{sec:obs_an}

Apart from XMM-1 and the DDT observation, we also use eight short \swift pointed observations obtained in 2025 to monitor the source state, together with the long-term \swift light curve to trace the spin-period evolution \footnote{\url{https://swift.gsfc.nasa.gov/results/transients/weak/3A0114p650.orbit.lc.txt}}. The observation log is given in Table~\ref{tab:obslog}. \xmmnewton EPIC cameras cover the $\sim$0.15--15 keV energy range \citep{2001A&A...365L..27T,2001A&A...365L..18S}. The \textit{Neil Gehrels Swift Observatory} operates over $\sim$15--150~keV and provides long-term monitoring of bright Galactic sources such as \object{4U\,0114+65} \citep{2005SSRv..120..143B}.

\begin{table}
\caption{Observation log for \object{4U\,0114+65}.}
\label{tab:obslog}
\centering
\begin{tabular}{lcc}
\hline\hline
Instrument & Date (UTC) & Duration (ks) \\
\hline
Swift/BAT & 2025-06-21 11:18:57 & 1.642 \\
Swift/BAT & 2025-06-21 12:51:57 & 1.698 \\
Swift/BAT & 2025-06-21 14:25:57 & 1.575 \\
Swift/BAT & 2025-06-21 15:58:57 & 1.635 \\
Swift/BAT & 2025-06-21 17:32:56 & 1.639 \\
Swift/BAT & 2025-06-21 19:05:57 & 1.699 \\
Swift/BAT & 2025-06-21 20:38:57 & 1.578 \\
Swift/BAT & 2025-07-24 18:03:56 & 1.010 \\
\hline
XMM-Newton/EPIC & 2015-08-21 20:25:52 & 49 \\
XMM-Newton/EPIC & 2025-08-03 11:15:01 & 21 \\
\hline
\end{tabular}

\tablefoot{
Swift/BAT ObsIDs, in the order listed in the table: 00015874104, 00015874105, 00015874106, 00015874107, 00015874108, 00015874109, 00015874110, and 00015874111.
The XMM-Newton ObsIDs are 0764650101 (PI: Torrej\'on; GO) for the 2015 observation, and 0971190901 and 0971190601 (PI: Schartel; DDT) for the 2025 observation.
Durations correspond to the on-source exposures reported by the mission archives, converted to ks.
}
\end{table}

The \textit{XMM-Newton}/EPIC data were reduced with the \textit{Science Analysis Software} (SAS, version 22.1.0). For the spectral analysis, the spectra from the three EPIC cameras, MOS1, MOS2, and pn, were combined into a single spectrum using the SAS task \texttt{epicspeccombine}. The combined spectra were subsequently grouped using \texttt{specgroup}, requiring a minimum of 25 counts per bin and an oversampling factor of 3. The grouped spectra were then loaded into the Interactive Spectral Interpretation System package (\texttt{ISIS})\footnote{\url{https://space.mit.edu/cxc/isis/}}, with no additional rebinning applied, for its analysis.

All quoted uncertainties correspond to the 90\% confidence level for one parameter of interest. Emission-line identifications were supported using the \textsc{atomdb} database\footnote{\url{http://www.atomdb.org/}} and \texttt{XSTAR} \citep{2021Atoms...9...12M}.

Orbital modulations were analyzed with the Python package \texttt{xraybinaryorbit} \citep{Sanjurjo-Ferrin2024}\footnote{\url{https://xragua.github.io/xraybinaryorbit/}}. Candidate emission lines were searched for with our publicly available \texttt{ISIS}-based tool \texttt{BLiSS}\footnote{\url{https://github.com/xragua/laex_detect_lines_isis}}. This automatic line-detection pipeline is designed to be continuum-independent, and can therefore be applied to spectra with different continuum shapes. Combined with \texttt{ISIS}, the code iterates over the detected candidate lines and retains those that produce an improvement in the reduced $\chi^{2}$ above the chosen threshold.

To detect short-timescale features in the light curves, such as dips and spikes, we used the Python package \texttt{dipspeaks} \citep{dipspeaks_zenodo}\footnote{\url{https://github.com/xragua/dipspeaks}}. The method generates synthetic light curves that reproduce the noise properties of the observed data, allowing the expected distribution of non-astrophysical fluctuations to be characterized. Significant deviations from this distribution in the real light curves are then flagged as likely astrophysical events.

\section{Timing analysis}
\label{sec:timing}

\begin{figure*}[ht]
    \centering
    \includegraphics[width=1\textwidth]{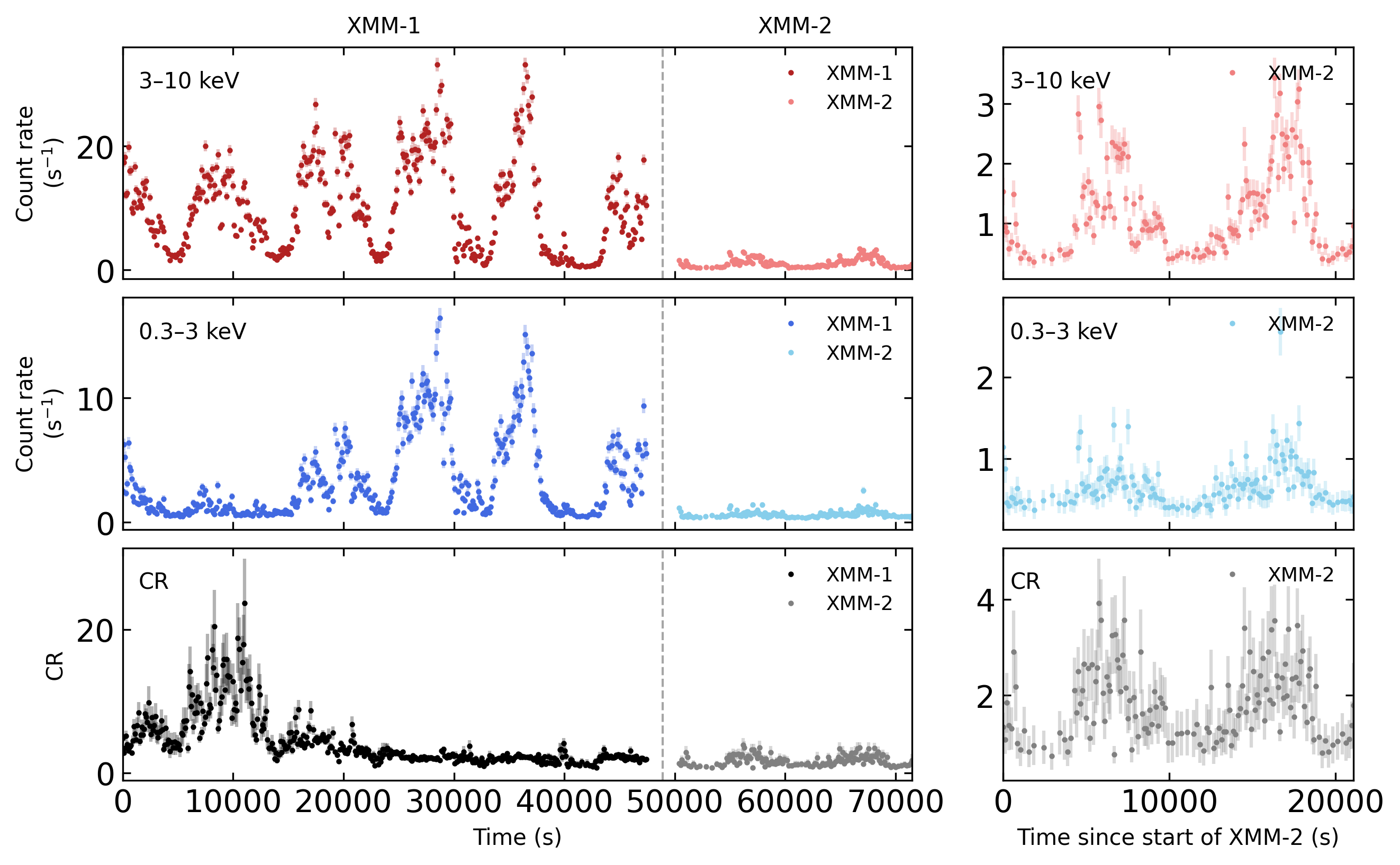}
    \caption{
    Upper left panel: combined XMM-1 and XMM-2 light curves in the hard X-ray band (3--10 keV). 
    Middle left panel: combined XMM-1 and XMM-2 light curves in the soft X-ray band (0.3--3 keV). 
    Upper right panel:  XMM-2 light curve in the hard X-ray band.
    Middle right panel: XMM-2 light curve in the soft X-ray band.
    Bottom left panel: combined XMM-1 and XMM-2 color ratio (CR, hard energy band divided by the soft energy band).  Bottom right panel: XMM-2 CR.
    }
    \label{fig:lc}
\end{figure*}

In 4U~0114+65, the NS modulation associated with the spin period can be clearly identified by eye when observed with \xmmnewton (see Fig. \ref{fig:lc}, top panels), even when the pulse is no longer detected with \swift (top right panels). Hereafter, we refer to the high-emission intervals as pulse peaks, corresponding to phases in which the NS hot spot faces the observer, and to the low-emission intervals as pulse valleys, when the hot spot is directed away from the observer. The most evident difference between XMM-1 and XMM-2 is the strong suppression of the X-ray emission during XMM-2, both in the soft and hard energy bands, as shown in Fig.~\ref{fig:lc}.

The complex temporal variability of 4U~0114+65 is clearly illustrated in Fig.~\ref{fig:lc}. The XMM-1 observation,  shown in  red for the hard band (3--10 keV) and in blue for the soft band(0.3--3 keV), can be divided into two intervals. The first interval is characterized by a relatively high color ratio (CR, hard energy band divided by the soft energy band) and an enhanced absorption column ($N_{\rm H}$; see Fig.~\ref{fig:lc}, lower left panel). In \citet{2017A&A...606A.145S}, this feature was tentatively associated with the passage of a co-rotating interaction region. We caution, however, that the winds of OB supergiants are highly structured, and that other large-scale features, such as an accretion wake, a tidal stream, or dense filaments and clumps, could produce a similar signature; the present data do not allow us to discriminate among these possibilities. The second interval displays a lower CR and no comparable absorption enhancement. Towards the end of the observation, the valley preceding the final pulse lasts significantly longer than expected. This feature was interpreted as an off-state, possibly associated with a reorganization of the accretion flow between the two magnetic poles, as described in \citet{2017A&A...606A.145S}.

 The CR also shows an interesting behavior. During the first interval of XMM-1 and throughout XMM-2, the CR is modulated with the pulse, whereas during the second interval of XMM-1 it becomes irregular, with  the modulation apparently recovering in the last pulse peak, after the off-state. This change in the CR behavior might also be related to a reorganization of the accretion flow between the magnetic poles, although testing this hypothesis would require an in-depth analysis beyond the scope of this work.

In addition to these large-scale variations, individual pulses show substantial pulse-to-pulse variability. The unusually long spin period of the NS allows us to inspect the intra-pulse structure in detail and to study changes from one pulse to the next.

\begin{figure*}[ht]
    \centering
    \includegraphics[width=1\textwidth]{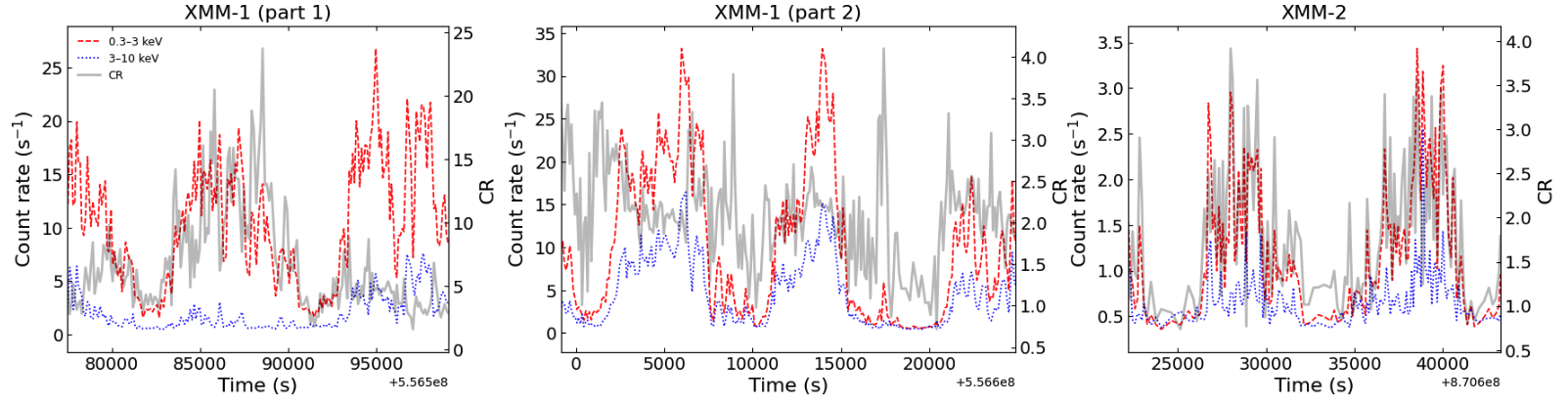}
    \caption{X-ray light curves and CR evolution for the two 
    \textit{XMM-Newton} observations. The XMM-1 observation is divided 
    into its two temporal intervals (left and central panels; see text), 
    while the complete XMM-2 observation is shown in the right panel. The 
    0.3--3~keV and 3--10~keV count rates are shown as red dashed and blue 
    dotted lines, respectively (left-hand vertical axes). The CR is shown 
    in gray (right-hand vertical axes). The vertical scales are adjusted 
    independently in each panel to emphasize the temporal evolution within 
    each interval; apparent amplitudes should therefore not be compared 
    directly between panels.}
    \label{fig:cr}
\end{figure*}

\subsection{Evolution of the NS spin from the \swift data}
 
\begin{figure}[ht]
    \centering
    \includegraphics[width=0.99\columnwidth]{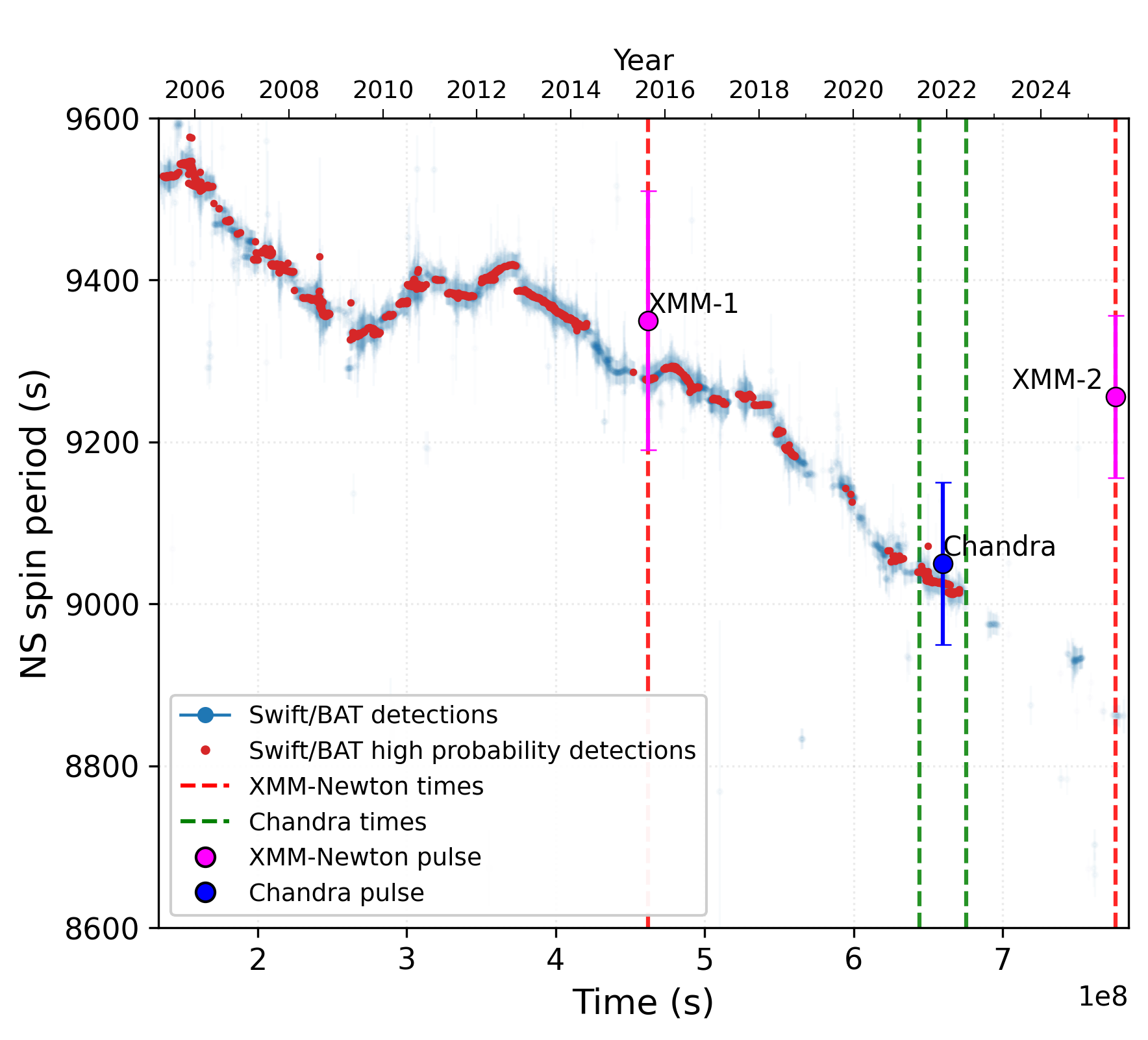}
    \caption{
    Time evolution of the NS spin period derived from the \swift light curve using a sliding-window Lomb--Scargle analysis. 
    In each window, we retained the candidate solution with the lowest false-alarm probability and considered only significant detections with a signal-to-noise ratio > 5 and a false-alarm probability $<10^{-5}$, meaning that such a peak has a probability lower than $10^{-5}$ of arising from random noise. These reliable detections are highlighted in red.
    The blue points and error bars show all measured periods and their uncertainties without signal-to-noise or false alarm probability cuts. Their opacity is proportional to their signal-to-noise ratio. 
    The red and green dashed lines indicate the times of the \textit{XMM-Newton} and \textit{Chandra} pointed observations, respectively. 
    The pulse periods obtained from these pointed observations are over-plotted for comparison.
    }
    \label{fig:sb_evolution}
\end{figure}

Over the years, 4U~0114+65 has shown a predominantly spin-up evolution, punctuated by occasional spin-down episodes. To trace the temporal evolution of the NS spin period, we performed a sliding-window timing analysis of the \textit{Swift}/BAT light curve using the \texttt{period\_sliding\_window} routine of the \texttt{xraybinaryorbit} package \citep{laex_2024_14263006}. The \textit{Swift}/BAT transient-monitor light curve is not uniformly 
sampled: each data point corresponds to a single-orbit snapshot, with 
a median separation between consecutive points of $\sim$5~ks 
(16th--84th percentiles: $\sim$1--11~ks). Specifically, a Lomb--Scargle periodogram was computed in successive 1000-bin intervals, shifted by 100 bins, probing trial periods in the 5000--15000\,s range. For each window, we selected the candidate period associated with the lowest false-alarm probability. We then retained only statistically significant detections, requiring a positive signal-to-noise ratio of at least 5 and a false-alarm probability below 10$^{-5}$. The resulting spin-period evolution is shown in Fig.~\ref{fig:sb_evolution}.

The spin periods measured in the pointed observations are consistent with the long-term \swift evolution. During XMM-1, the observed NS spin period was $\sim$9350\,s, while during the 2021--2022 \chandra observations it was close to $\sim$9100\,s. These values agree with the periods inferred from the \swift monitoring at similar epochs.

When we requested \xmmnewton DDT to observe this source, our expectations were based on the previously observed \swift trend. We anticipated that, if the NS spin period were detectable, it would be of the order of 8.9\,ks. However, the data revealed the opposite behavior: the spin period was found to be longer than that measured in the 2021--2022 \chandra observations, returning instead to values comparable to those observed by \xmmnewton in 2015.

The last robust period measurement before XMM-2—here defined as having a signal-to-noise ratio $>5$ and a false-alarm probability $<10^{-5}$—was obtained with \swift in 2021, roughly four years prior to the observation of XMM-2, and indicated a spin period of $\sim 9$\,ks. This result is further supported by its simultaneity with a series of pointed \chandra observations, which independently confirm this spin period. We note that the exact epoch of the last reliable detection depends on the adopted selection cuts, so that different thresholds may lead to a somewhat different set of accepted periods. 

When the modulation is detected, its properties are instead indicative of a spin-down of the NS spin pulse. In a similar analysis, \citet{2017ApJ...844...16H} found that the long-term spin evolution of 4U~0114+65 is highly variable, with alternating spin-up and spin-down or random-walk epochs. They interpreted the strong spin-up episodes as possible evidence for the transient formation of an accretion disc, while the spin-down or random-walk intervals were associated with phases dominated by direct wind accretion.

\subsection{Spikes in the light curves}

On timescales of seconds to a few kiloseconds, the X-ray light curves of HMXBs display short-lived features in the form of dips and spikes. The latter are characterized by their duration and prominence, defined as the increase in count rate above the local baseline. The unusually long spin period of the NS is an advantage here, as individual pulses last $\sim$9~ks, so that short-timescale features can be inspected within each pulse instead of being diluted, as would be the case in faster pulsars. In \citet{2025A&A...694A.192S}, these events were associated with instabilities operating at two characteristic locations of the accretion flow: short spikes with Rayleigh--Taylor instabilities (RTIs) at the magnetospheric boundary, and longer spikes with variability close to the Bondi radius. Since the soft-band emission is strongly suppressed in XMM-2, we restrict this analysis to the hard-band (3--10~keV) light curves of both observations.
The detection of such features poses two challenges: distinguishing genuine astrophysical events from statistical fluctuations, and doing so while retaining sufficient time resolution. To address them, we used the public algorithm \texttt{dipspeaks} \citep{dipspeaks_zenodo}, which implements a semi-supervised anomaly-detection scheme based on  autoencoders.

Autoencoders are neural networks trained to reconstruct their input; once trained, events that are poorly reconstructed (i.e., with a large reconstruction error) can be flagged as anomalies \citep{2020arXiv200305991B}. In our case, the algorithm proceeds as follows. First, synthetic light curves are generated as random time series sharing the statistical properties (noise distribution and sampling) of the real data but containing no astrophysical signal. Candidate spikes detected in these synthetic light curves therefore represent the population of spurious, noise-driven events, and are used to train the autoencoder. Spikes detected in the real light curve that deviate significantly from this noise population are retained as likely astrophysical events. In addition, we computed the signal-to-noise ratio (S/N) of each candidate from the mean uncertainty of the light curve around the event, and candidates with S/N $> 5$ were also accepted. Full details of the method are given in \citet{2025A&A...694A.192S} and in the \texttt{dipspeaks} documentation.

\begin{figure}
\centering
\includegraphics[width=1\columnwidth]{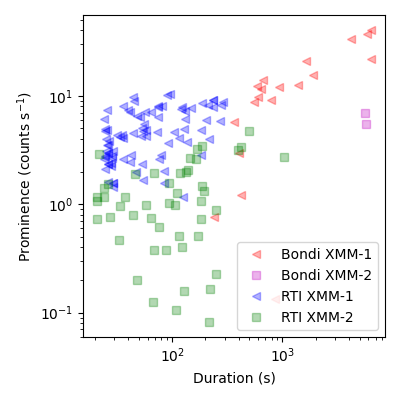}
\caption{
Spikes detected in XMM-1 and XMM-2. Short-duration spikes are tentatively associated with RTIs at the magnetospheric boundary, while longer-duration spikes are associated with variability close to the Bondi radius. Prominence is defined as the increase in count rate above the local baseline.}
\label{Fig:spikes}
\end{figure}

For each spike, we define the prominence as the increase in count rate above the local baseline.
After selecting the significant spikes, we applied a Gaussian mixture model clustering analysis, also implemented in \texttt{dipspeaks}. Unlike \textit{k}-means, which assumes approximately spherical clusters, Gaussian mixtures allow for more general cluster morphologies, including elliptical distributions and components with different sizes \citep{Reynolds2009}. In this case, the clustering yields a silhouette score above 0.6 in both observations, indicating the presence of two well-separated groups. The results are shown in Fig.~\ref{Fig:spikes}. The duration and prominence of the two spike populations are consistent with the physical interpretation proposed by \cite{2025A&A...694A.192S}.

\begin{table}
\caption{Summary of the spike clusters identified in XMM-1 and XMM-2.}
\label{tab:spike_clusters}

\begin{tabular}{llcc}
\hline
Obs. & Type & Prominence & Duration (s) \\
\hline
XMM-1 & Short & $4^{+4}_{-2}$ & $50^{+80}_{-20}$ \\ 
XMM-1 & Long & $12^{+13}_{-7}$ & $800^{+4000}_{-300}$ \\ 
XMM-2 & Short & $1.2^{+1.5}_{-0.8}$ & $120^{+100}_{-80}$ \\ 
XMM-2 & Long & $6.2 \pm 0.5$ & $5700\pm 40$ \\
\hline
\end{tabular}
\tablefoot{The listed values correspond to the median prominence and duration of each cluster. The lower and upper uncertainties are given by the 16th and 84th percentiles, respectively. Clusters are labelled as short and long according to their median duration.}
\end{table}

\section{Spectral analysis}
\label{sec:spectra}

We performed three complementary spectral analyses. First, we fitted the phase-averaged spectra extracted from the full exposure of XMM-1 and XMM-2, in order to characterize the global spectral properties of the source in XMM-1 and XMM-2. Second, we carried out spin-phase-resolved spectroscopy to identify which continuum components are modulated with the NS rotation and to trace the spectral changes as the accretion-powered hot spot rotates into and out of the line of sight. Finally, we searched for Fe\,K$\alpha$ emission-line signatures in the individual spectra.

To facilitate comparison with previous results, we used the same spectral model as in \citet{2025A&A...694A.192S}. The continuum was described with the bulk-motion Comptonization model, \texttt{bmc}. This analytical model describes the Comptonization of soft seed photons by matter undergoing relativistic bulk motion \citep{1997ApJ...487..834T}. Its main parameters are the characteristic blackbody temperature of the soft photon source, the spectral energy index $\alpha$, and an illumination parameter, $\log(A)$, that describes the fractional illumination of the bulk-motion flow by the thermal photon source.

The \texttt{bmc} illumination parameter was fixed at $\log(A)=1$, corresponding to a Comptonized fraction $f=A/(1+A)\simeq0.91$. This value was retained, as in our previous analysis \citep{2025A&A...694A.192S}, to reduce model degeneracy and to allow a direct comparison with the \textit{Chandra} results. Leaving both $\alpha$ and $\log(A)$ free resulted in poorly constrained and strongly degenerate \texttt{bmc} parameters.

Photoelectric absorption was modeled with the Tuebingen--Boulder absorption model, \texttt{Tbnew}. This model computes the X-ray absorption cross section as the sum of the contributions from the gas-phase ISM, grain-phase ISM, and molecules in the ISM \citep{2000ApJ...542..914W}. Emission lines were modeled with Gaussian components when required.

The soft excess was fitted with a \texttt{bbody} component. When left free, the temperature and absorption column of this component were consistently found in the ranges 0.04--0.07 keV and $(0.5$--$0.8)\times10^{22}$ cm$^{-2}$, respectively. We therefore fixed the temperature to the most frequent value, 0.06 keV, and fixed the absorption column ($N_{\rm H}$), to the interstellar value towards the source, $0.8\times10^{22}$ cm$^{-2}$, inferred from the optical reddening $E(B-V)$ listed in Table~\ref{4u_parameters}. The complete model is described by

\begin{equation} 
\label{eq:model}
\begin{aligned}
F(E) & = \exp[-N_{\rm H}\sigma(E)]\times \left[{\rm BMC}(E)+G_{\rm Fe\,K\alpha}(E) \right] \\
& \quad + \exp[-N_{\rm H}^{\rm ISM}\sigma(E)]\times {\rm BBody}(E).
\end{aligned}
\end{equation}

The radius of the region emitting the seed soft photons, which are subsequently Comptonized, can be estimated by assuming that this region radiates as a blackbody of area $\pi R_{W}^{2}$ \citep{2004A&A...423..301T}:

\begin{equation}
\label{radius}
R_{W}=0.6\sqrt{L_{34}}(kT)^{-2} \ {\rm km},
\end{equation}
where $L_{34}$ is the luminosity of the \texttt{bmc} component in units of $10^{34}$ erg s$^{-1}$, and $kT$ is the temperature of the seed soft photons.

\begin{figure*}[ht]
    \centering
    \includegraphics[width=1\textwidth]{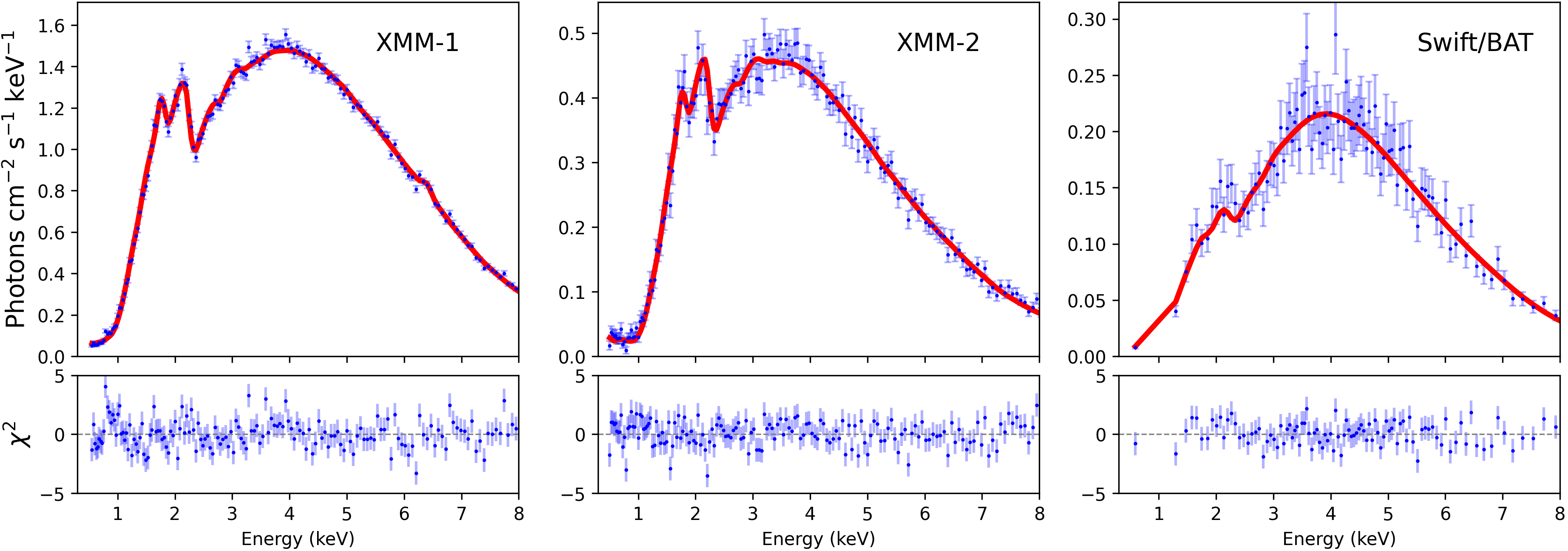}
    \caption{
    Average spectra of XMM-1 and XMM-2. Upper panels show the observed spectra in blue and the best-fit model in red. Lower panels show the corresponding residuals in units of $\chi^2$. The model corresponds to Eq.~\ref{eq:model}, and the best-fit parameters are listed in Table~\ref{tab:average_spectra}.
    }
    \label{fig:av_spectra}
\end{figure*}

The best-fit parameters obtained for the phase-averaged spectra are listed in Table~\ref{tab:average_spectra}. The corresponding spectra, best-fit models, and residuals are shown in Fig.~\ref{fig:av_spectra}. 

\begin{figure*}[ht]
    \centering
    \includegraphics[width=1\textwidth]{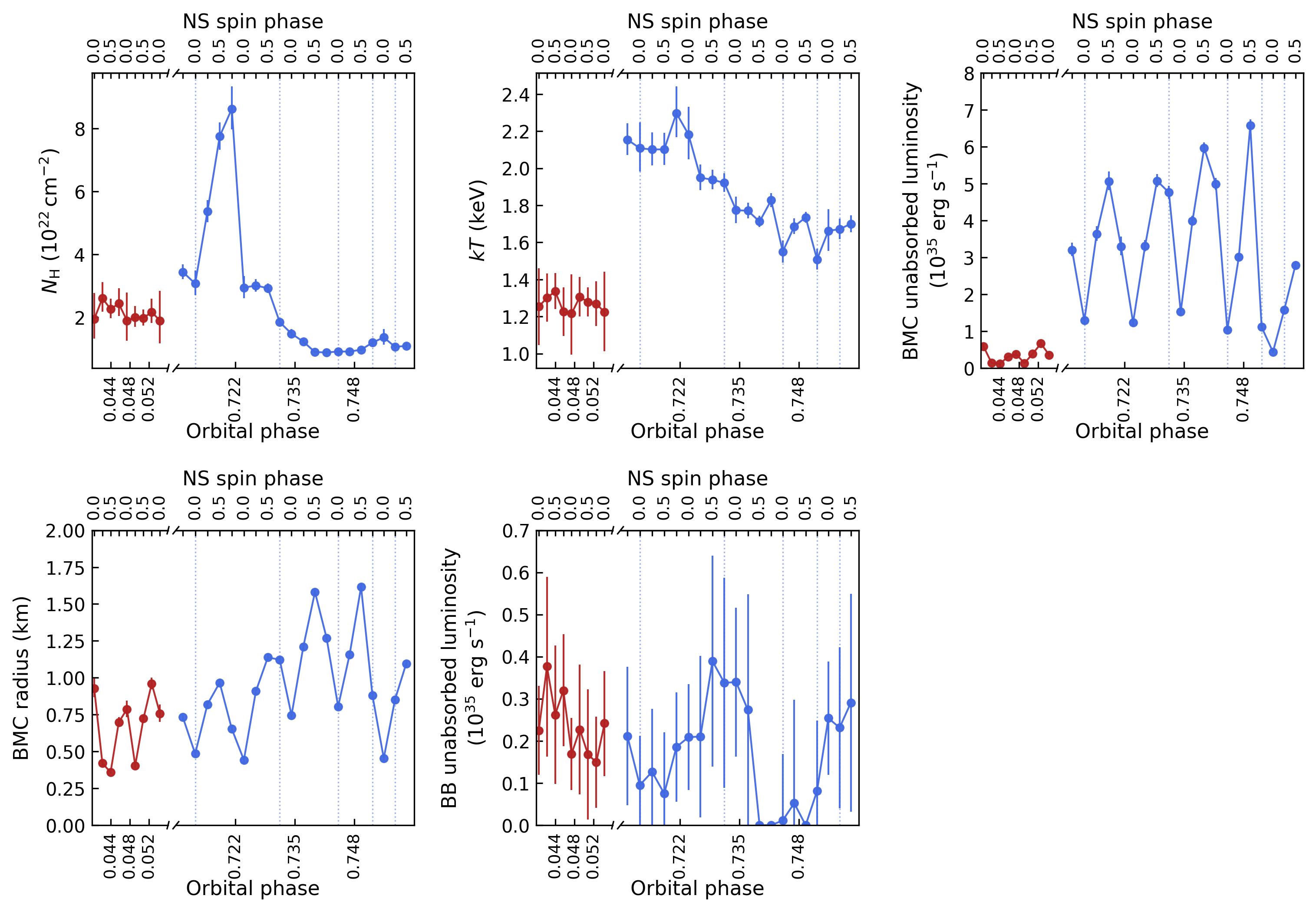}
    \caption{
    Orbital-phase evolution of selected continuum parameters and inferred quantities for the pointed observations of 4U~0114$+$65. The phase bins were chosen to match the broad NS-spin-phase intervals defined in Sect.~4 (0.12--0.37, 0.37--0.62, 0.62--0.87, and 0.87--0.12). The five panels,  read from the upper left to the lower right, show the equivalent hydrogen column density $N_{\rm H}$, the seed-photon temperature, the unabsorbed \texttt{bmc} luminosity, the \texttt{bmc} radius, and the unabsorbed \texttt{bbody} luminosity. XMM-2 is shown in red and XMM-1 in blue. To facilitate the comparison between observations obtained at different orbital phases, the x-axis is displayed in separated phase windows. Vertical dotted lines indicate the orbital phases where Fe~K$\alpha$ signatures are found in XMM-1.
    }
    \label{fig:average_spectra_ev}
\end{figure*}

We then performed spin-phase-resolved spectroscopy to investigate the spectral variability along the NS spin cycle. For the phase-resolved analysis, the \texttt{bmc} parameter $\alpha$ was fixed to the best-fitting value obtained from the corresponding average spectrum, since it could not be independently constrained in all phase intervals. This assumes that the spectral slope remains approximately constant within each observation. The \texttt{bmc} normalization was left free because variations in the observed emission site (hot spot on the NS surface) are expected over the NS rotational cycle. We note, however, that because $\alpha$ and the \texttt{bmc} normalization are partially degenerate, fixing $\alpha$ may cause part of the spectral variability to be absorbed by the normalization. Therefore, the phase dependence of the \texttt{bmc} normalization should be interpreted with this caveat in mind.

We used two complementary approaches. First, spectra were extracted in four broad spin-phase intervals: 0.12--0.37, 0.37--0.62, 0.62--0.87, and 0.87--0.12. These intervals correspond approximately to the rise of the pulse, the phase in which the hot spot is directed towards the observer, the decay of the pulse, and the phase in which the hot spot is directed away from the observer, respectively. This broad spin-resolved analysis allows us to identify the main spectral changes associated with the NS spin cycle and through time. The resulting best-fit parameters are reported in Tables~\ref{tab:phase_resolved_appendix1} and \ref{tab:phase_resolved_appendix2}. The orbital-phase evolution of the continuum parameters and derived quantities is shown in Fig.~\ref{fig:average_spectra_ev}. Note that x axis points to orbital modulation while the upper x axis refers to the NS spin pulse, i.e. the analysis shows the evolution with time corresponding to bins of the NS spin phase intervals. 

For XMM-1, shown in blue, the first panel reveals an interval of  enhanced $N_{\rm H}$, accompanied by a CR increase (see  Fig.~\ref{fig:lc}, lower left panel). In \citet{2017A&A...606A.145S}, this feature was tentatively associated with the passage of a CIR. We caution, however, that the winds of OB supergiants are highly structured, and that other large-scale features, such as an  accretion wake, a tidal stream, or dense filaments and clumps, could produce a similar signature. The present data do not allow us to discriminate among these possibilities. In addition, during the final pulse peak of XMM-1, the modulation of the \texttt{bmc} component appears damped, coinciding with the off-state identified in the light curve (left panels of Fig.~\ref{fig:lc}).

The \texttt{bmc} component is clearly modulated with the NS spin. We computed the pulse fraction of this component as

\begin{equation}
{\rm PF} = \frac{F_{\rm max}-F_{\rm min}}{F_{\rm max}+F_{\rm min}},
\end{equation}
where $F_{\rm max}$ and $F_{\rm min}$ are the maximum and minimum unabsorbed fluxes of the \texttt{bmc} component along the NS spin cycle, respectively. In both XMM-1 and XMM-2, the observed pulse fraction is $\sim$0.7.

\begin{figure*}[ht]
    \centering
    \includegraphics[width=1\textwidth]{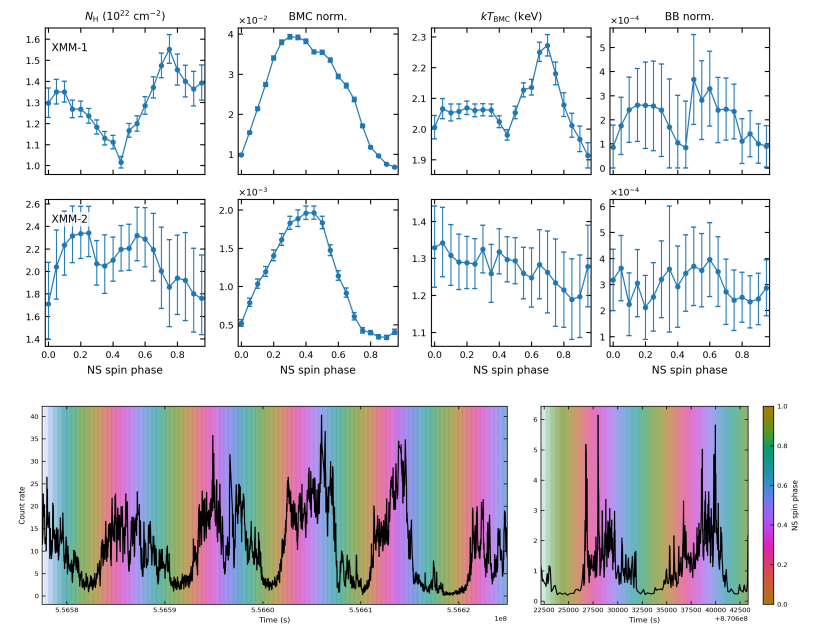}
    \caption{
    NS-spin-folded spectral analysis of 4U~0114$+$65.
    The upper and middle rows show the evolution of the best-fitting continuum parameters as a function of NS folded spin phase for XMM-1 and XMM-2, respectively. From left to right, the panels display the equivalent hydrogen column density, the \texttt{bmc} normalization, the \texttt{bmc} temperature, and the blackbody normalization. The lower panels show the time intervals selected for the extraction of the NS-spin-folded spectra, overlaid on the light curves. Each color denotes a different spin-phase bin; intervals with the same color were combined to obtain the final spectral bins.
    }
    \label{fig:spin_folded_spectra}
\end{figure*}

Second, to further inspect the detailed shape of the spectral modulation with the NS spin, we performed a higher-resolution NS-spin-folded spectral analysis. In this case, the spin cycle was divided into narrower bins of width $\Delta\phi_{\rm spin}=0.05$. Since individual bins do not provide sufficient signal-to-noise ratios for independent spectral fitting, spectra corresponding to the same spin-phase interval were combined over the full duration of each observation. This resulted in 20 spectra for both XMM-1 and XMM-2, which were fitted with the same continuum model used in the phase-averaged analysis. The corresponding best-fit parameters are listed in Tables~\ref{tab:ns_spin_folded_xmm1} and \ref{tab:ns_spin_folded_xmm2}. The results of the high-resolution spin-folded analysis are shown in Fig.~\ref{fig:spin_folded_spectra}.

The Fe\,K$\alpha$ emission line is expected to trace dense and relatively cold material, such as clumps or weakly ionized regions of the stellar wind. In 4U~0114$+$65, this line is much weaker than is typically observed in HMXBs. The companion star in 4U~0114$+$65, a B1~Ia supergiant, is among the coolest donors in supergiant X-ray binaries \citep{2015A&A...579A.111K,2020A&A...643A...9E}. With an effective temperature of $T_{\rm eff}=24$~kK, the source lies close to the bistability jump \citep[see the lower panel of Fig.~3 in][]{1999A&A...350..181V}. This jump is caused by a sudden change in the ionization balance of Fe (\textsc{iii}, \textsc{iv}), which is a major contributor to the radiative acceleration of the wind. \citet{2017A&A...606A.145S} suggested that this may affect the efficiency of clump formation and/or destruction, concluding that 4U~0114$+$65 has the typical dense wind of a supergiant star, but with a much lower degree of clumping.

We searched for Fe\,K$\alpha$ signatures by adding an emission-line component to each spectrum and evaluating the resulting improvement in $\chi^2$. To estimate the significance of these features, we used the \texttt{fakeit} command in \texttt{ISIS} to generate 1000 continuum-only simulated spectra, which were then searched for spurious emission lines in the 5--8 keV range. A line was considered a Fe\,K$\alpha$ detection when its inclusion improved the fit above the 99\% significance threshold derived from the simulated spectra. In addition, we required the inferred line flux to be inconsistent with zero.

\begin{table}
\caption{Best-fit Gaussian line parameters for the tentative Fe K$\alpha$ signatures identified in XMM-1.}
\label{tab:gaussian_lines}
\centering
\begin{tabular}{cccc}
\hline\hline
Bin & Center & Area & $\sigma$ \\
    & (keV) & ($10^{-5}\,\mathrm{ph\,cm^{-2}\,s^{-1}}$) & ($\mathrm{eV}$) \\
\hline
2  & $6.41^{+0.04}_{-0.06}$ & $7 \pm 5$ & $<5$ \\
9  & $6.35 \pm 0.01$         & $9 \pm 7$ & $<5$ \\
15 & $6.35 \pm 0.01$         & $8 \pm 6$ & $<5$ \\
18 & $6.44^{+0.01}_{-0.07}$ & $5 \pm 3$ & $<5$ \\
20 & $6.40 \pm 0.05$         & $4 \pm 3$ & $<5$ \\
\hline
\end{tabular}
\end{table}

Under these criteria, we found no significant Fe~K$\alpha$ detections. However, five tentative Fe~K$\alpha$ signatures were identified in XMM-1, while no comparable signatures were found in XMM-2. The corresponding line parameters are listed in Table~\ref{tab:gaussian_lines}, and the orbital phases at which these signatures are found are marked by vertical blue lines in Fig.~\ref{fig:average_spectra_ev}. In XMM-1, the Fe\,K$\alpha$ signatures appear mainly during phases in which the hot spot is directed away from the observer. This is consistent with the line being produced in a more extended region, such as clumps or the smooth stellar wind, and becoming diluted when the direct emission from the hot spot dominates the observed spectrum. This behavior may also be partly driven by contrast effects, since the Fe\,K$\alpha$ line becomes easier to detect when the continuum level decreases.

\section{Discussion}
The results presented above indicate that the disappearance of the pulse in the Swift/BAT monitoring is not caused by a complete loss of the spin modulation. Instead, the sensitive XMM-2 observation shows that the pulse persists at lower luminosity and lower absolute pulsed flux. In the following, we discuss whether this change is primarily driven by variations in the accretion-powered hot-spot emission, by changes in the donor wind, or by a reduced efficiency of plasma entry through the NS magnetosphere.

\subsection{The hot-spot emission and accretion}
The X-ray emission in this system is produced by accretion onto the NS magnetic poles, with the \texttt{bmc} component providing the dominant contribution to the continuum. The spin-resolved spectral analysis shows that both the \texttt{bmc} normalization and the unabsorbed luminosity vary with the NS spin phase in both observations (see Fig.~\ref{fig:average_spectra_ev} for the evolution through the observation in bins corresponding to NS spin intervals and Tables \ref{tab:phase_resolved_appendix1} and \ref{tab:phase_resolved_appendix2} and Fig.~\ref{fig:spin_folded_spectra} for the spin folded spectra and Tables \ref{tab:ns_spin_folded_xmm1} and \ref{tab:ns_spin_folded_xmm2}). They reach their maximum values when the hot spot is expected to be most directly visible, and decrease when it rotates away from the line of sight.

We use two complementary spin-resolved approaches. First, broad spin-phase intervals are used to trace the overall evolution of the continuum parameters along the NS spin and time. Second, the spectra folded into narrower spin-phase bins are used to inspect the detailed shape of the modulation and to compare the behavior of XMM-1 and XMM-2 at higher NS spin phase resolution.

Interestingly, the overall spectral shape, and even the \texttt{bmc} pulse fraction, are very similar in both observations (Fig.~\ref{fig:av_spectra} and Table~\ref{tab:average_spectra}). The main difference is that the unabsorbed luminosity is approximately one order of magnitude higher in XMM-1 than in XMM-2, with luminosities of $3.3 \times 10^{35}$ erg s$^{-1}$ and $3.1 \times 10^{34}$ erg s$^{-1}$, respectively. This difference is also evident in the soft and hard X-ray light curves (Fig.~\ref{fig:lc}), where the emission in XMM-2 is strongly suppressed relative to XMM-1.

The luminosity decrease is mainly concentrated in the \texttt{bmc} component, whose unabsorbed flux is about a factor of $\sim 9$ higher in XMM-1 than in XMM-2. The unabsorbed flux of the \texttt{bbody} component is also higher in XMM-1, but only by a factor of $\sim 2$. This indicates that the luminosity drop appears to be primarily associated with the suppression of the accretion-powered hot-spot emission.

The higher-resolution NS-spin-folded spectral analysis (Fig.~\ref{fig:spin_folded_spectra}) confirms that the \texttt{bmc} component is clearly modulated with the NS rotation. 

The modulation appears broader in XMM-1, whereas in XMM-2 it is narrower and occurs at a lower absolute normalization. The \texttt{bmc} temperature is lower in XMM-2 than in XMM-1, showing in XMM-1 a mild increase towards later spin phases, 
possibly indicating that the seed-photon emitting region is not strictly uniform. 
This may point to an extended or structured hot spot, although degeneracies 
among the continuum parameters cannot be excluded. 

In XMM-1, the minimum of $N_{\rm H}$ in the NS-spin folded spectra (see Fig. \ref{fig:spin_folded_spectra}) occurs close to the spin phases where the \texttt{bmc} normalization reaches its maximum. This may indicate that, when the accretion-powered hot spot is most directly visible, the material along the line of sight is more strongly photoionized by the enhanced X-ray emission. In this case, the lower fitted $N_{\rm H}$ is due to a lower effective opacity of the intervening material. This interpretation is consistent with the higher luminosity during XMM-1.

In the fainter XMM-2 observation, the weaker ionizing continuum may lead to a less clearly modulated effective absorption column. Nevertheless, a shallow and non-significant decrease in $N_{\rm H}$ may be present between NS spin phases $\sim0.2$ and $\sim0.3$, approximately coincident with the \texttt{bmc} maximum. 

In slowly rotating wind-fed pulsars, accretion onto the NS can be mediated by plasma entry through Rayleigh--Taylor instabilities at the magnetospheric boundary. This regime has been studied for several decades \citep[see][]{1976ApJ...207..914A,1977ApJ...215..897E}. At moderate X-ray luminosities, below $\sim 4 \times 10^{36}$ erg s$^{-1}$, a hot convective quasi-spherical shell may form above the NS magnetosphere. In this settling-accretion regime, the radial velocity of the wind plasma is lower than the free-fall velocity, and the mass entry rate through the magnetosphere is regulated by the efficiency of Compton and radiative cooling.

In \citet{2025A&A...694A.192S}, spikes were classified into short and long events and associated with different accretion instabilities, including instabilities close to the Bondi radius and Rayleigh--Taylor instabilities close to the magnetosphere. That classification was not based on fixed thresholds in duration or prominence, but was obtained using a \texttt{Gaussian Mixture} clustering algorithm. Applying the same approach here, we find that the short-spike population, which we tentatively associate with RTI-driven accretion events, is more prominent in XMM-1 than in XMM-2 and has shorter characteristic durations (Table~\ref{tab:spike_clusters}). This behaviour is qualitatively consistent with the expected dependence of the instability timescale on luminosity, since the characteristic free-fall timescale at the Alfvén radius can be written as
\begin{equation}
t_A \sim \frac{R_A^{3/2}}{(GM)^{1/2}}.
\end{equation}

Assuming that the spectral shape does not change strongly during these short events, the spike prominence can be used as an observational proxy for the excess effective accretion rate onto the NS. This approximation is particularly useful in the 3--10 keV band, where the radiation is less affected by absorption in the circumstellar material. Under this assumption,
\begin{equation}
\Delta \dot{M}_{\rm acc,eff} \propto P_i \, \tau_i,
\end{equation}
where $P_i$ and $\tau_i$ are the prominence, defined as an increase in count rate above the local baseline, and duration of the $i$-th spike, respectively. We therefore define an observational proxy for the time-averaged contribution of short accretion events as
\begin{equation}
Q_{\rm RTI} \equiv \frac{1}{T_{\rm obs}} \sum_i P_i \,\tau_i,
\end{equation}
where $T_{\rm obs}$ is the total exposure time.

We find that this relative spike contribution is approximately 3--4 times larger in XMM-1 than in XMM-2. This increase is significant, but it is still smaller than the overall factor of $\sim 9$ difference observed in the \texttt{bmc} flux. This suggests that the luminosity difference between the two observations is not entirely driven by RTI accretion alone. Instead, XMM-1 likely includes an additional underlying accretion component, producing a brighter quasi-persistent continuum that is strongly suppressed in XMM-2.

\subsection{Absorption column}

The hydrogen column density along the line of sight is computed by integrating the hydrogen number density between the compact object and the observer:
\begin{equation}
N_{\rm H} = \int_{l_0}^{\infty} n_{\rm H}(l)\, dl ,
\end{equation}
where $n_{\rm H}$ is the hydrogen number density and $l$ denotes the coordinate along the line of sight. As a first-order approximation, we assume that $n_{\rm H}$ scales linearly with the wind density. Under the assumption of a stationary, spherically symmetric stellar wind, the radial density profile is given by
\begin{equation}
\rho(r) = \frac{\dot{M}}{4\pi r^2 v(r)},
\end{equation}
with the wind velocity described by the standard $\beta$-law (see footnote Table \ref{4u_parameters}).

where $\dot{M}$ is the donor stellar mass-loss rate, $v_\infty$ is the terminal wind velocity, and $\beta$ characterizes the wind acceleration profile.

To constrain the system parameters, we fitted the measured $N_{\rm H}$ values using the \texttt{fit\_nh\_ps} routine from the \texttt{xraybinaryorbit} package \citep{Sanjurjo-Ferrin2024}. This routine uses particle swarm optimization (PSO) to minimize the difference between the observed and model-predicted column densities. PSO is a gradient-free global optimization technique in which a population of candidate solutions explores the parameter space. This method is particularly well suited to non-linear problems with correlated parameters, where gradient-based approaches may converge slowly or become trapped in local minima.

As expected for an inclined and eccentric system, the absorption column is higher in XMM-2 than in XMM-1, since XMM-2 was obtained closer to superior conjunction. We fitted the $N_{\rm H}$ values from both observations independently, using their corresponding orbital phases, in order to assess whether the circumstellar environment changed between epochs. Specifically, for XMM-2 we adopted the phase-averaged value, whereas for XMM-1 we used only the values obtained between orbital phases 0.73--0.75, thereby excluding the initial portion of the observation, which is affected by enhanced absorption.

As a result, we obtained $\dot{M} = (7.4 \pm 0.2)\times10^{-7}\,M_{\odot}\,\mathrm{yr^{-1}}$ for XMM-1 and $\dot{M} = (5.6 \pm 0.3)\times10^{-7}\,M_{\odot}\,\mathrm{yr^{-1}}$ for XMM-2. The value inferred for XMM-2 is lower than that of XMM-1, although both remain of the same order of magnitude. We stress that the quoted statistical uncertainties do not account for the substantially larger systematic uncertainties associated with the adopted wind model and its parameters. Under the assumptions of our simplified smooth and spherically symmetric wind model, this suggests that no major change in the global donor wind is required between the two epochs. In this context, the weakening of the \texttt{bmc} component may not be driven exclusively by a lower amount of circumstellar material available for accretion, but could also reflect changes in the structure or radiative properties of the accretion flow near the NS.

We caution that the absolute values of $\dot{M}$ derived here depend critically on the adopted wind velocity law  parameters, $v_\infty$ and $\beta$, which are poorly known in general, vary from source to source, and may be affected by photoionization feedback from the NS. Our comparison should therefore be regarded as differential: assuming that the global wind parameters did not vary between the two epochs, the similarity of the inferred mass-loss rates indicates that the observed $N_{\rm H}$ does not require a substantially different amount of circumstellar material in XMM-2 with respect to XMM-1.

\subsection{Toward centrifugal inhibition?}

The observed properties of 4U~0114+65 can be interpreted within the framework of the quasi-spherical settling accretion model for slowly rotating, magnetized NSs developed by \citet{2012MNRAS.420..216S} and further discussed by \citet{2014EPJWC..6402001S} (see also \citealt{2015ARep...59..645S,Shakura:2018fQ} for reviews). This wind-fed accretion regime is expected in systems with moderate X-ray luminosities, below $\simeq 4\times 10^{36}$~erg~s$^{-1}$, where a hot, convective, quasi-spherical shell forms above the NS magnetosphere. In this picture, the rate at which plasma enters the magnetosphere is regulated by the efficiency of plasma cooling, while the shell mediates the transfer of angular momentum to and from the rotating NS.

Under these conditions, the equilibrium spin period of the NS is mainly determined by the wind velocity captured at the Bondi radius, the orbital period, and the NS magnetic field, conveniently parametrized through the dipole magnetic moment $\mu = B_0R_{\rm NS}^3/2$:
\begin{equation}
P_{\rm eq} (s) \approx 1000\,{\rm s}\,
\mu_{30}^{12/11}
\left(\frac{P_b}{10\,\mathrm{d}}\right)
\dot M_{16}^{-4/11}
v_8^4 \, 
\label{4u_e:Peq}
\end{equation}
\citep[]{2012MNRAS.420..216S}.

Here, $\mu_{30}\equiv\mu/(10^{30}\,\mathrm{G\,cm}^3)$, $\dot M_{16}\equiv \dot M/(10^{16}\,\mathrm{g\,s}^{-1})$, and $v_8\equiv v_{\rm rel}/(10^8\,\mathrm{cm\,s}^{-1})$. For 4U~0114+65 $v_8\sim0.5$--0.7 (Table~\ref{4u_parameters}). Given the observed pulse period, Eq.~(\ref{4u_e:Peq}) points to a large magnetic field, of the order of $\mu_{30}\sim100$, that is, in the magnetar range. 

The long-term spin evolution of 4U~0114+65 may further suggest that the source is approaching a regime in which accretion becomes progressively less efficient. Within the quasi-spherical settling accretion scenario, this does not necessarily imply the onset of a fully developed propeller state. Instead, it may indicate an intermediate situation in which centrifugal effects begin to play a non-negligible role in regulating plasma entry through the magnetosphere.

The relevant quantities are the magnetospheric, or Alfv\'en, radius, $R_{\rm A}$, and the corotation radius, $R_{\rm co}$. For slowly rotating, wind-fed NSs in the settling accretion regime, the Alfv\'en radius can be approximated as
\begin{equation}
R_{\rm A} \simeq 1.4\times10^{9}\,\mu_{30}^{6/11}\,\dot{M}_{16}^{-2/11}\ {\rm cm},
\end{equation}
as derived for the settling accretion regime by \citet{2012MNRAS.420..216S}; see also \citealt{Shakura:2018fQ}), where the different dependence on $\mu$ and $\dot{M}$ with respect to the standard Alfvén radius ($R_{\rm A}\propto\mu^{4/7}\dot{M}^{-2/7}$) arises from the pressure balance with the hot quasi-static shell rather than with a free-falling flow, while the corotation radius is
\begin{equation}
R_{\rm co} =
\left(
\frac{G M_{\rm NS} P_{\rm spin}^{2}}{4\pi^{2}}
\right)^{1/3}.
\end{equation}
A fully developed centrifugal barrier is expected only when $R_{\rm A} \gtrsim R_{\rm co}$. However, the ratio $R_{\rm A}/R_{\rm co}$ provides a useful diagnostic of how close the system may be to centrifugal inhibition.

\begin{figure}[ht]
    \centering
    \includegraphics[width=1\columnwidth]{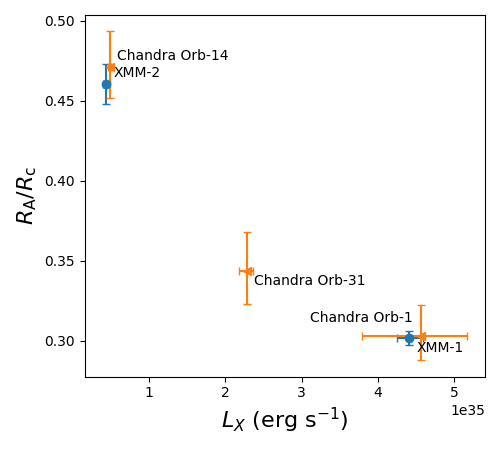}
    \caption{
    Ratio between the Alfv\'en radius and the corotation radius, $R_{\rm A}/R_{\rm co}$, as a function of X-ray unabsorbed luminosity for the different observations of 4U~0114+65. The points correspond to the \xmmnewton, \chandra, and \swift\ measurements. Error bars account for the uncertainties in both luminosity and spin period.
    }
    \label{fig:ratio_lum}
\end{figure}

For the adopted magnetic moment and the luminosity-derived accretion rates, this ratio increases from approximately 0.3 to 0.5 over the range sampled by the available observations, as shown in Fig.~\ref{fig:ratio_lum}. Although the source remains below the canonical propeller threshold, the increase in $R_{\rm A}/R_{\rm co}$ suggests that it may be evolving towards a regime in which centrifugal effects become increasingly relevant.

This trend is physically meaningful. A secular spin-up reduces the corotation radius,
\begin{equation}
R_{\rm co}\propto P_{\rm spin}^{2/3},
\end{equation}
whereas a decrease in accretion rate increases the Alfv\'en radius,
\begin{equation}
R_{\rm A}\propto \dot{M}^{-2/11}.
\end{equation}
Therefore, a source that becomes fainter while maintaining, or having recently undergone, a spin-up episode will naturally evolve towards larger values of $R_{\rm A}/R_{\rm co}$.

In such a scenario, accretion is not completely halted, but plasma entry through the magnetospheric boundary becomes progressively less efficient and possibly more intermittent. This picture is qualitatively consistent with the cyclical accretion scenario previously proposed for the 2021 \chandra\ observation of 4U~0114+65. In that scenario, matter accumulates above the magnetosphere during faint intervals and is later accreted more efficiently once Compton cooling enhances the Rayleigh--Taylor instability.

Within the interpretation proposed here, the secular spin evolution may amplify this cyclical behaviour. During relatively bright states, the accretion rate may be high enough to sustain efficient torque transfer and continued spin-up. As the spin period decreases, $R_{\rm co}$ also decreases. If the accretion rate subsequently drops, $R_{\rm A}$ increases, so that the ratio $R_{\rm A}/R_{\rm co}$ becomes larger. Centrifugal effects could then partially inhibit the inflow, reducing the overall accretion efficiency and leading to lower luminosities and weaker or less easily detectable pulsations.

This hypothesis is qualitatively consistent with the pulsation history shown in Fig.~\ref{fig:sb_evolution}. In 2015, pulsations were clearly detected with a spin period of $\sim 9.4$~ks. Subsequent detections suggest that the period continued to decrease, until the first extended interval without significant pulse detections appeared around 2017. Pulsations were detected again in 2020 and 2021, with a period of $\sim 9.1$~ks, when the source also displayed the cyclical accretion behavior discussed above. In 2025, pulsations were detected again with \xmmnewton, but with a longer period of $\sim 9.3$~ks. These pulsations were weaker than in previous pointed observations and remained undetected in the \swift\ monitoring data.

This behavior may indicate that, after the spin-up phase inferred from the 2015--2021 detections, the accretion flow became less efficient at maintaining a positive torque on the NS. The system may therefore have entered a phase dominated by weaker or less coherent accretion torques, allowing the spin period to drift back towards longer values, closer to those expected in the settling-accretion regime.

We stress, however, that the present values of $R_{\rm A}/R_{\rm co}$ do not support a fully developed propeller regime. Rather, they point to an intermediate situation in which the source may be approaching centrifugal gating, with partial suppression of accretion and a possible coupling between luminosity changes and torque evolution.

It is also possible to relate the drastic spin-down of the source observed in Fig.~\ref{fig:sb_evolution} (between $\sim 8950$ s at the last \textit{Swift}/BAT point and the XMM-2 value of $\sim 9250$ s), $\dot P_{*,sd}\sim 10^{-4}$ s s$^{-1}$, to the maximum possible spin-down rate during the settling accretion regime. As shown in \cite{2012MNRAS.420..216S}, the maximum spin-up and spin-down rates of slowly rotating accreting X-ray pulsars are expected in the so-called strong-coupling regime between the accreting matter and rotating NS magnetosphere. In this case, turbulent motions of the accreting matter above the magnetosphere can lead to the growth of the toroidal magnetic-field component up to a maximum value comparable to the poloidal component, $B_t\sim B_p$. The maximum spin-down torque is then $I\dot \omega_*\lesssim K_2\mu^2/R_A^3$, where $I$ is the NS moment of inertia, $\omega_\ast=2\pi/P_\mathrm{spin}$, and $K_2\approx 7.56$; the equality is reached in the propeller regime, when accretion is centrifugally inhibited. Taking $\mu\sim 10^{32}$ G cm$^3$, $R_A\sim 1.7 \times 10^{10}$ cm, and $P_\mathrm{spin}=9050$ s (see Table~\ref{4u_parameters}), we obtain $\dot{P}_{\rm theor}\lesssim 1.5\times 10^{-4}$ s s$^{-1}$, close to the observed value. This estimate suggests that the low-accretion state of the source might have been caught when the strong-coupling regime was established. As shown in \cite{2012MNRAS.420..216S}, the strong-coupling regime between the accreting matter and rotating NS magnetosphere is characterized by rapid change of spin-up/spin-down episodes. Therefore, we may have evidenced such a transition in the  XMM-observations. Here we note that, in such a slowly rotating pulsar as 4U~0114+65, with the corotation radius comparable to the Bondi radius, fully centrifugally inhibited accretion is hardly possible, as part of the gravitationally captured matter can couple to the magnetosphere much closer to the magnetic poles, where centrifugal effects are negligible.

\section{Summary and conclusions}

We have analyzed a 2025 \textit{XMM-Newton} Director's Discretionary Time observation of 4U~0114$+$65, and compared it with a previous \textit{XMM-Newton} observation obtained in 2015, with the aim of investigating the origin of the apparent disappearance of the long X-ray pulse in the \textit{Swift}/BAT monitoring. Our main results can be summarized as follows.

\begin{enumerate}
    \item The 2025 \xmmnewton observation reveals the NS spin with a period of about 9.3~ks, despite the fact that pulsations remained undetected in the long-term \textit{Swift}/BAT data. This indicates that the pulse did not vanish completely, but rather became too weak to be significantly detected in the BAT data.\\
       
    \item The overall spectral shape is broadly similar in the 2015 and 2025 \xmmnewton observations, but the source was much fainter in 2025. The unabsorbed luminosity decreased by about one order of magnitude, mainly because of a strong suppression of the bulk-motion Comptonization component, while the soft blackbody-like excess changed much less strongly. The radius inferred for the Comptonized component remains compatible with emission from a hot spot on the NS surface. The phase dependence of the continuum parameters is also consistent with the changing visibility of the accretion hot spot along the NS spin cycle.\\

    \item The absorption column is higher in XMM-2, as expected from its orbital phase, but the inferred wind parameters remain of the same order of magnitude. This suggests that the luminosity decrease does not require a major global change in the donor wind. Instead, the suppression of the \texttt{bmc} component is more likely related to changes in the accretion flow close to the NS.\\

    \item The short-timescale spike analysis shows that the contribution of short events, tentatively associated with Rayleigh--Taylor instabilities near the magnetosphere, is larger in XMM-1 than in XMM-2. However, this difference is smaller than the overall decrease in the \texttt{bmc} flux, suggesting that the luminosity drop is not driven by short accretion events alone, but also by a reduction of the underlying quasi-persistent accretion-powered emission.\\

    \item The long-term evolution of the source can be naturally interpreted in terms of a progressive increase in the ratio between the magnetospheric and corotation radii. In this picture, in the strong coupling regime, 4U~0114$+$65 may be approaching a regime of partial centrifugal inhibition, in which plasma entry through the magnetosphere becomes less efficient and more intermittent, without reaching a fully developed propeller state.\\

    \item The rapid spin-down measured between the last \textit{Swift}/BAT detection and XMM-2, $\dot{P}_{*,\mathrm{sd}} \sim 10^{-4}$~s\,s$^{-1}$, is close to the maximum spin-down rate expected in the strong-coupling regime of quasi-spherical settling accretion, in which the toroidal magnetic-field component grows up to values comparable to the poloidal one, $B_{\mathrm{t}} \sim B_{\mathrm{p}}$. Although this interpretation remains tentative, if confirmed it would represent the first observational evidence of a transition to the strong-coupling regime in an accreting X-ray pulsar.

\end{enumerate}

\section*{Data availability}
The \textit{XMM-Newton} and \textit{Swift}/BAT data used in this work are publicly available through the corresponding mission archives. 

\begin{acknowledgements}
GSF, JMT, JJRR, and JPV acknowledge the financial support from the MICIU/AEI/10.13039/501100011033 with funding from the European Union (FEDER). Project (NewAthena24-UA), reference PID2024-155779OB-C33.
This work is based on observations obtained with XMM-Newton through Director’s Discretionary Time (DDT). We thank Norbert Schartel for granting the DDT observations. We acknowledge the constructive criticism of the referee whose comments improved the content of the paper.
\end{acknowledgements}

\bibliographystyle{aa}
\bibliography{example.bib} 

@ARTICLE{1977IAUC.3144....2D,
   author = {{Dower}, R. and {Kelley}, R. and {Margon}, B. and {Bradt}, H.
	},
    title = "{2S 0114+650}",
  journal = {\iaucirc},
     year = 1977,
    month = nov,
   volume = 3144,
   adsurl = {http://adsabs.harvard.edu/abs/1977IAUC.3144....2D}
}

@ARTICLE{1999ApJ...513L..45L,
   author = {{Li}, X.-D. and {van den Heuvel}, E.~P.~J.},
    title = "{Could 2S 0114+650 Be a Magnetar?}",
  journal = {\apjl},
   eprint = {astro-ph/9901084},
     year = 1999,
    month = mar,
   volume = 513,
    pages = {L45-L48},
      doi = {10.1086/311904},
   adsurl = {http://cdsads.u-strasbg.fr/abs/1999ApJ...513L..45L}
}

@ARTICLE{2000ApJ...536..450H,
   author = {{Hall}, T.~A. and {Finley}, J.~P. and {Corbet}, R.~H.~D. and 
	{Thomas}, R.~C.},
    title = "{RXTE Observations of the X-Ray Binary 2S 0114+650}",
  journal = {\apj},
     year = 2000,
    month = jun,
   volume = 536,
    pages = {450-454},
      doi = {10.1086/308924},
   adsurl = {http://cdsads.u-strasbg.fr/abs/2000ApJ...536..450H}
}

@ARTICLE{2006MNRAS.367.1457F,
   author = {{Farrell}, S.~A. and {Sood}, R.~K. and {O'Neill}, P.~M.},
    title = "{Super-orbital period in the high-mass X-ray binary 2S 0114+650}",
  journal = {\mnras},
   eprint = {astro-ph/0502008},
     year = 2006,
    month = apr,
   volume = 367,
    pages = {1457-1462},
      doi = {10.1111/j.1365-2966.2006.10150.x},
   adsurl = {http://adsabs.harvard.edu/abs/2006MNRAS.367.1457F}
}

@ARTICLE{2011MNRAS.413.1083W,
   author = {{Wang}, W.},
    title = "{Long-term hard X-ray monitoring of 2S 0114+65 with INTEGRAL/IBIS}",
  journal = {\mnras},
archivePrefix = "arXiv",
   eprint = {1012.3211},
 primaryClass = "astro-ph.HE",
     year = 2011,
    month = may,
   volume = 413,
    pages = {1083-1098},
      doi = {10.1111/j.1365-2966.2010.18192.x},
   adsurl = {http://adsabs.harvard.edu/abs/2011MNRAS.413.1083W}
}

@ARTICLE{1985ApJ...299..839C,
   author = {{Crampton}, D. and {Hutchings}, J.~B. and {Cowley}, A.~P.},
    title = "{The supergiant X-ray binary system 2S 0114 + 650}",
  journal = {\apj},
     year = 1985,
    month = dec,
   volume = 299,
    pages = {839-844},
      doi = {10.1086/163750},
   adsurl = {http://cdsads.u-strasbg.fr/abs/1985ApJ...299..839C}
}

@ARTICLE{2012MNRAS.420..216S,
   author = {{Shakura}, N. and {Postnov}, K. and {Kochetkova}, A. and {Hjalmarsdotter}, L.
	},
    title = "{Theory of quasi-spherical accretion in X-ray pulsars}",
  journal = {\mnras},
archivePrefix = "arXiv",
   eprint = {1110.3701},
 primaryClass = "astro-ph.HE",
     year = 2012,
    month = feb,
   volume = 420,
    pages = {216-236},
      doi = {10.1111/j.1365-2966.2011.20026.x},
   adsurl = {http://adsabs.harvard.edu/abs/2012MNRAS.420..216S}
}

@INPROCEEDINGS{2014EPJWC..6402001S,
   author = {{Shakura}, N.~I. and {Postnov}, K.~A. and {Kochetkova}, A.~Y. and 
	{Hjalmarsdotter}, L.},
    title = "{Theory of wind accretion}",
booktitle = {European Physical Journal Web of Conferences},
     year = 2014,
   series = {European Physical Journal Web of Conferences},
   volume = 64,
archivePrefix = "arXiv",
   eprint = {1307.3029},
 primaryClass = "astro-ph.HE",
    month = jan,
      eid = {02001},
    pages = {02001},
      doi = {10.1051/epjconf/20136402001},
   adsurl = {http://adsabs.harvard.edu/abs/2014EPJWC..6402001S}
}

@ARTICLE{2004A&A...423..301T,
   author = {{Torrej{\'o}n}, J.~M. and {Kreykenbohm}, I. and {Orr}, A. and 
	{Titarchuk}, L. and {Negueruela}, I.},
    title = "{Evidence for a Neutron Star in the non-pulsating massive X-ray binary 4U2206+54}",
  journal = {\aap},
   eprint = {astro-ph/0405182},
     year = 2004,
    month = aug,
   volume = 423,
    pages = {301-309},
      doi = {10.1051/0004-6361:20035743},
   adsurl = {http://adsabs.harvard.edu/abs/2004A%26A...423..301T}
}

@ARTICLE{Reig,
   author = {{Reig}, P. and {Chakrabarty}, D. and {Coe}, M.~J. and {Fabregat}, J. and 
	{Negueruela}, I. and {Prince}, T.~A. and {Roche}, P. and {Steele}, I.~A.
	},
    title = "{Astrophysical parameters of the massive X-ray binary 2S 0114+650.}",
  journal = {\aap},
     year = 1996,
    month = jul,
   volume = 311,
    pages = {879-888},
   adsurl = {http://cdsads.u-strasbg.fr/abs/1996A%26A...311..879R}
}

@ARTICLE{2006A&A...458..513K,
   author = {{Koenigsberger}, G. and {Georgiev}, L. and {Moreno}, E. and 
	{Richer}, M.~G. and {Toledano}, O. and {Canalizo}, G. and {Arrieta}, A.
	},
    title = "{The X-ray binary 2S0114+650=LSI+65 010. A slow pulsar or tidally-induced pulsations?}",
  journal = {\aap},
   eprint = {astro-ph/0608226},
     year = 2006,
    month = nov,
   volume = 458,
    pages = {513-522},
      doi = {10.1051/0004-6361:20065305},
   adsurl = {http://adsabs.harvard.edu/abs/2006A%26A...458..513K}
}

@ARTICLE{2001A&A...365L..18S,
   author = {{Str{\"u}der}, L. and {Briel}, U. and {Dennerl}, K. and {Hartmann}, R. and 
	{Kendziorra}, E. and {Meidinger}, N. and {Pfeffermann}, E. and 
	{Reppin}, C. and {Aschenbach}, B. and {Bornemann}, W. and {Br{\"a}uninger}, H. and 
	{Burkert}, W. and {Elender}, M. and {Freyberg}, M. and {Haberl}, F. and 
	{Hartner}, G. and {Heuschmann}, F. and {Hippmann}, H. and {Kastelic}, E. and 
	{Kemmer}, S. and {Kettenring}, G. and {Kink}, W. and {Krause}, N. and 
	{M{\"u}ller}, S. and {Oppitz}, A. and {Pietsch}, W. and {Popp}, M. and 
	{Predehl}, P. and {Read}, A. and {Stephan}, K.~H. and {St{\"o}tter}, D. and 
	{Tr{\"u}mper}, J. and {Holl}, P. and {Kemmer}, J. and {Soltau}, H. and 
	{St{\"o}tter}, R. and {Weber}, U. and {Weichert}, U. and {von Zanthier}, C. and 
	{Carathanassis}, D. and {Lutz}, G. and {Richter}, R.~H. and 
	{Solc}, P. and {B{\"o}ttcher}, H. and {Kuster}, M. and {Staubert}, R. and 
	{Abbey}, A. and {Holland}, A. and {Turner}, M. and {Balasini}, M. and 
	{Bignami}, G.~F. and {La Palombara}, N. and {Villa}, G. and 
	{Buttler}, W. and {Gianini}, F. and {Lain{\'e}}, R. and {Lumb}, D. and 
	{Dhez}, P.},
    title = "{The European Photon Imaging Camera on XMM-Newton: The pn-CCD camera}",
  journal = {\aap},
     year = 2001,
    month = jan,
   volume = 365,
    pages = {L18-L26},
      doi = {10.1051/0004-6361:20000066},
   adsurl = {http://adsabs.harvard.edu/abs/2001A%26A...365L..18S}
}

@ARTICLE{2013ARep...57..287I,
   author = {{Ikhsanov}, N.~R. and {Beskrovnaya}, N.~G.},
    title = "{The spin-down mechanism of the X-ray pulsar 4U 2206+54}",
  journal = {Astronomy Reports},
archivePrefix = "arXiv",
   eprint = {1211.6314},
 primaryClass = "astro-ph.HE",
     year = 2013,
    month = apr,
   volume = 57,
    pages = {287-293},
      doi = {10.1134/S1063772913030013},
   adsurl = {http://adsabs.harvard.edu/abs/2013ARep...57..287I}
}

@ARTICLE{2000ApJ...542..914W,
   author = {{Wilms}, J. and {Allen}, A. and {McCray}, R.},
    title = "{On the Absorption of X-Rays in the Interstellar Medium}",
  journal = {\apj},
   eprint = {astro-ph/0008425},
     year = 2000,
    month = oct,
   volume = 542,
    pages = {914-924},
      doi = {10.1086/317016},
   adsurl = {http://adsabs.harvard.edu/abs/2000ApJ...542..914W}
}

@ARTICLE{1997ApJ...487..834T,
   author = {{Titarchuk}, L. and {Mastichiadis}, A. and {Kylafis}, N.~D.},
    title = "{X-Ray Spectral Formation in a Converging Fluid Flow: Spherical Accretion into Black Holes}",
  journal = {\apj},
   eprint = {astro-ph/9702092},
     year = 1997,
    month = oct,
   volume = 487,
    pages = {834-846},
   adsurl = {http://cdsads.u-strasbg.fr/abs/1997ApJ...487..834T}
}

@ARTICLE{2015ARep...59..645S,
   author = {{Shakura}, N.~I. and {Postnov}, K.~A. and {Kochetkova}, A.~Y. and 
	{Hjalmarsdotter}, L. and {Sidoli}, L. and {Paizis}, A.},
    title = "{Wind accretion: Theory and observations}",
  journal = {Astronomy Reports},
archivePrefix = "arXiv",
   eprint = {1407.3163},
 primaryClass = "astro-ph.HE",
     year = 2015,
    month = jul,
   volume = 59,
    pages = {645-655},
      doi = {10.1134/S1063772915070112},
   adsurl = {http://adsabs.harvard.edu/abs/2015ARep...59..645S}
}

@ARTICLE{1992A&A...262L..25F,
   author = {{Finley}, J.~P. and {Belloni}, T. and {Cassinelli}, J.~P.},
    title = "{Periodic outbursts in the peculiar X-ray binary 2S 0114+65}",
  journal = {\aap},
     year = 1992,
    month = sep,
   volume = 262,
    pages = {L25-L28},
   adsurl = {http://adsabs.harvard.edu/abs/1992A%26A...262L..25F}
}

@ARTICLE{2005A&A...436L..31B,
   author = {{Bonning}, E.~W. and {Falanga}, M.},
    title = "{INTEGRAL high energy observations of 2S 0114+65}",
  journal = {\aap},
   eprint = {astro-ph/0504633},
     year = 2005,
    month = jun,
   volume = 436,
    pages = {L31-L34},
      doi = {10.1051/0004-6361:200500117},
   adsurl = {http://adsabs.harvard.edu/abs/2005A%26A...436L..31B}
}

@ARTICLE{2015MNRAS.454.4467P,
   author = {{Pradhan}, P. and {Paul}, B. and {Paul}, B.~C. and {Bozzo}, E. and 
	{Belloni}, T.~M.},
    title = "{Is 4U 0114+65 an eclipsing HMXB?}",
  journal = {\mnras},
archivePrefix = "arXiv",
   eprint = {1509.08706},
 primaryClass = "astro-ph.HE",
     year = 2015,
    month = dec,
   volume = 454,
    pages = {4467-4475},
      doi = {10.1093/mnras/stv2276},
   adsurl = {http://adsabs.harvard.edu/abs/2015MNRAS.454.4467P}
}

@ARTICLE{2012MNRAS.425..595R,
   author = {{Reig}, P. and {Torrej{\'o}n}, J.~M. and {Blay}, P.},
    title = "{Accreting magnetars: a new type of high-mass X-ray binaries?}",
  journal = {\mnras},
archivePrefix = "arXiv",
   eprint = {1203.1490},
 primaryClass = "astro-ph.HE",
     year = 2012,
    month = sep,
   volume = 425,
    pages = {595-604},
      doi = {10.1111/j.1365-2966.2012.21509.x},
   adsurl = {http://adsabs.harvard.edu/abs/2012MNRAS.425..595R}
}

@ARTICLE{2015A&A...579A.111K,
   author = {{Krti{\v c}ka}, J. and {Kub{\'a}t}, J. and {Krti{\v c}kov{\'a}}, I.
	},
    title = "{X-ray irradiation of the winds in binaries with massive components}",
  journal = {\aap},
archivePrefix = "arXiv",
   eprint = {1505.03411},
 primaryClass = "astro-ph.SR",
     year = 2015,
    month = jul,
   volume = 579,
      eid = {A111},
    pages = {A111},
      doi = {10.1051/0004-6361/201525637},
   adsurl = {http://adsabs.harvard.edu/abs/2015A%26A...579A.111K}
}

@ARTICLE{1999A&A...350..181V,
   author = {{Vink}, J.~S. and {de Koter}, A. and {Lamers}, H.~J.~G.~L.~M.
	},
    title = "{On the nature of the bi-stability jump in the winds of early-type supergiants}",
  journal = {\aap},
   eprint = {astro-ph/9908196},
     year = 1999,
    month = oct,
   volume = 350,
    pages = {181-196},
   adsurl = {http://adsabs.harvard.edu/abs/1999A%26A...350..181V}
}

@ARTICLE{2008ApJ...678..408L,
   author = {{Lobel}, A. and {Blomme}, R.},
    title = "{Modeling Ultraviolet Wind Line Variability in Massive Hot Stars}",
  journal = {\apj},
archivePrefix = "arXiv",
   eprint = {0712.3804},
     year = 2008,
    month = may,
   volume = 678,
      eid = {408-430},
    pages = {408-430},
      doi = {10.1086/529129},
   adsurl = {http://adsabs.harvard.edu/abs/2008ApJ...678..408L}
}

@BOOK{1999isw..book.....L,
   author = {{Lamers}, H.~J.~G.~L.~M. and {Cassinelli}, J.~P.},
    title = "{Introduction to Stellar Winds}",
booktitle = {Introduction to Stellar Winds, by Henny J.~G.~L.~M.~Lamers and Joseph P.~Cassinelli, pp.~452.~ISBN 0521593980.~Cambridge, UK: Cambridge University Press, June 1999.},
     year = 1999,
    month = jun,
    pages = {452},
   adsurl = {http://adsabs.harvard.edu/abs/1999isw..book.....L}
}

@ARTICLE{grun,
   author = {{Grundstrom}, E.~D. and {Blair}, J.~L. and {Gies}, D.~R. and 
	{Huang}, W. and {McSwain}, M.~V. and {Raghavan}, D. and {Riddle}, R.~L. and 
	{Subasavage}, J.~P. and {Wingert}, D.~W. and {Levine}, A.~M. and 
	{Remillard}, R.~A.},
    title = "{Joint H{$\alpha$} and X-Ray Observations of Massive X-Ray Binaries. I. The B Supergiant System LS I +65 010 = 2S 0114+650}",
  journal = {\apj},
   eprint = {astro-ph/0610898},
     year = 2007,
    month = feb,
   volume = 656,
    pages = {431-436},
      doi = {10.1086/510508},
   adsurl = {http://adsabs.harvard.edu/abs/2007ApJ...656..431G}
}

@ARTICLE{1992ApJ...392L...9D,
   author = {{Duncan}, R.~C. and {Thompson}, C.},
    title = "{Formation of very strongly magnetized neutron stars - Implications for gamma-ray bursts}",
  journal = {\apjl},
     year = 1992,
    month = jun,
   volume = 392,
    pages = {L9-L13},
      doi = {10.1086/186413},
   adsurl = {http://adsabs.harvard.edu/abs/1992ApJ...392L...9D}
}

@ARTICLE{2008ApJ...683.1031B,
   author = {{Bozzo}, E. and {Falanga}, M. and {Stella}, L.},
    title = "{Are There Magnetars in High-Mass X-Ray Binaries? The Case of Supergiant Fast X-Ray Transients}",
  journal = {\apj},
archivePrefix = "arXiv",
   eprint = {0805.1849},
     year = 2008,
    month = aug,
   volume = 683,
      eid = {1031-1044},
    pages = {1031-1044},
      doi = {10.1086/589990},
   adsurl = {http://adsabs.harvard.edu/abs/2008ApJ...683.1031B}
}

@ARTICLE{2006AdSpR..38.2779S,
   author = {{Sood}, R. and {Farrell}, S. and {O'Neill}, P. and {Manchanda}, R. and 
	{Ashok}, N.~M.},
    title = "{Evolution of the periodicities in 2S 0114+650}",
  journal = {Advances in Space Research},
   eprint = {astro-ph/0603008},
     year = 2006,
    month = jan,
   volume = 38,
    pages = {2779-2781},
      doi = {10.1016/j.asr.2006.02.063},
   adsurl = {http://adsabs.harvard.edu/abs/2006AdSpR..38.2779S}
}

@ARTICLE{2001A&A...378L..21O,
   author = {{Oskinova}, L.~M. and {Clarke}, D. and {Pollock}, A.~M.~T.},
    title = "{Rotationally modulated X-ray emission from the single O star {$\zeta$} Ophiuchi}",
  journal = {\aap},
     year = 2001,
    month = oct,
   volume = 378,
    pages = {L21-L24},
      doi = {10.1051/0004-6361:20011222},
   adsurl = {http://adsabs.harvard.edu/abs/2001A%26A...378L..21O}
}

@ARTICLE{2014MNRAS.441.2173M,
   author = {{Massa}, D. and {Oskinova}, L. and {Fullerton}, A.~W. and {Prinja}, R.~K. and
        {Bohlender}, D.~A. and {Morrison}, N.~D. and {Blake}, M. and
        {Pych}, W.},
    title = "{CIR modulation of the X-ray flux from the O7.5 III(n)((f)) star {$\xi$} Persei}",
  journal = {\mnras},
archivePrefix = "arXiv",
   eprint = {1403.5601},
 primaryClass = "astro-ph.SR",
     year = 2014,
    month = jul,
   volume = 441,
    pages = {2173-2180},
      doi = {10.1093/mnras/stu565},
   adsurl = {http://adsabs.harvard.edu/abs/2014MNRAS.441.2173M}}

@ARTICLE{2017A&A...606A.145S,
       author = {{Sanjurjo-Ferr{\'\i}n}, G. and {Torrej{\'o}n}, J.~M. and {Postnov}, K. and {Oskinova}, L. and {Rodes-Roca}, J.~J. and {Bernabeu}, G.},
        title = "{XMM-Newton spectroscopy of the accreting magnetar candidate 4U0114+65}",
      journal = {\aap},
         year = 2017,
        month = oct,
       volume = {606},
          eid = {A145},
        pages = {A145},
          doi = {10.1051/0004-6361/201630119},
archivePrefix = {arXiv},
       eprint = {1706.04907},
 primaryClass = {astro-ph.HE},
       adsurl = {https://ui.adsabs.harvard.edu/abs/2017A&A...606A.145S}
}

@ARTICLE{2023MNRAS.522.3271A,
       author = {{Abdallah}, Mohammed H. and {Samir}, Rasha M. and {Leahy}, Denis A. and {Shaker}, Ashraf A.},
        title = "{Nustar observation of the binary system 4U 0114 + 65}",
      journal = {\mnras},
         year = 2023,
        month = jul,
       volume = {522},
       number = {3},
        pages = {3271-3277},
          doi = {10.1093/mnras/stad1199},
archivePrefix = {arXiv},
       eprint = {2304.09295},
 primaryClass = {astro-ph.HE},
       adsurl = {https://ui.adsabs.harvard.edu/abs/2023MNRAS.522.3271A}
}

@ARTICLE{2017ApJ...844...16H,
       author = {{Hu}, Chin-Ping and {Chou}, Yi and {Ng}, C. -Y. and {Lin}, Lupin Chun-Che and {Yen}, David Chien-Chang},
        title = "{Evolution of Spin, Orbital, and Superorbital Modulations of 4U 0114+650}",
      journal = {\apj},
         year = 2017,
        month = jul,
       volume = {844},
       number = {1},
          eid = {16},
        pages = {16},
          doi = {10.3847/1538-4357/aa79a3},
archivePrefix = {arXiv},
       eprint = {1706.03902},
 primaryClass = {astro-ph.HE},
       adsurl = {https://ui.adsabs.harvard.edu/abs/2017ApJ...844...16H}
}

@ARTICLE{2020arXiv200305991B,
       author = {{Bank}, Dor and {Koenigstein}, Noam and {Giryes}, Raja},
        title = "{Autoencoders}",
      journal = {arXiv e-prints},
         year = 2020,
        month = mar,
          eid = {arXiv:2003.05991},
        pages = {arXiv:2003.05991},
          doi = {10.48550/arXiv.2003.05991},
archivePrefix = {arXiv},
       eprint = {2003.05991},
 primaryClass = {cs.LG},
       adsurl = {https://ui.adsabs.harvard.edu/abs/2020arXiv200305991B}
}

@Inbook{Reynolds2009,
author="Reynolds, Douglas",
editor="Li, Stan Z.
and Jain, Anil",
title="Gaussian Mixture Models",
bookTitle="Encyclopedia of Biometrics",
year="2009",
publisher="Springer US",
address="Boston, MA",
pages="659--663",
isbn="978-0-387-73003-5",
doi="10.1007/978-0-387-73003-5_196",
url="https://doi.org/10.1007/978-0-387-73003-5_196"
}

@article{Bailer-Jones_2021,
doi = {10.3847/1538-3881/abd806},
url = {https://dx.doi.org/10.3847/1538-3881/abd806},
year = {2021},
month = {feb},
publisher = {The American Astronomical Society},
volume = {161},
number = {3},
pages = {147},
author = {C. A. L. Bailer-Jones and J. Rybizki and M. Fouesneau and M. Demleitner and R. Andrae},
title = {Estimating Distances from Parallaxes. V. Geometric and Photogeometric Distances to 1.47 Billion Stars in Gaia Early Data Release 3},
journal = {The Astronomical Journal}
}

@ARTICLE{1976ApJ...207..914A,
       author = {{Arons}, J. and {Lea}, S.~M.},
        title = "{Accretion onto magnetized neutron stars: structure and interchange instability of a model magnetosphere.}",
      journal = {\apj},
         year = 1976,
        month = aug,
       volume = {207},
        pages = {914-936},
          doi = {10.1086/154562},
       adsurl = {https://ui.adsabs.harvard.edu/abs/1976ApJ...207..914A}
}

@ARTICLE{1977ApJ...215..897E,
       author = {{Elsner}, R.~F. and {Lamb}, F.~K.},
        title = "{Accretion by magnetic neutron stars. I. Magnetospheric structure and stability.}",
      journal = {\apj},
         year = 1977,
        month = aug,
       volume = {215},
        pages = {897-913},
          doi = {10.1086/155427},
       adsurl = {https://ui.adsabs.harvard.edu/abs/1977ApJ...215..897E}
}

@software{laex_2024_14263006,
  author       = {LAEX and
                  GracielaSanjurjoFerrin and
                  Matteo Bachetti},
  title        = {xragua/xraybinaryorbit: 1.0.1},
  month        = dec,
  year         = 2024,
  publisher    = {Zenodo},
  version      = {1.0.1},
  doi          = {10.5281/zenodo.14263006},
  url          = {https://doi.org/10.5281/zenodo.14263006},
  swhid        = {swh:1:dir:0ebde9a0ce78beb100baae922c088b2240face14
                   ;origin=https://doi.org/10.5281/zenodo.14216914;vi
                   sit=swh:1:snp:2bf45be53ad32a1faf69e424de0884497ac6
                   8349;anchor=swh:1:rel:0345a0a8f4bbf1279ac1f693fb72
                   2a4aa3e712f5;path=/
                  },
}

@article{Sanjurjo-Ferrin2024,
  doi = {10.21105/joss.07220},
  url = {https://doi.org/10.21105/joss.07220},
  year = {2024},
  publisher = {The Open Journal},
  volume = {9},
  number = {104},
  pages = {7220},
  author = {Graciela Sanjurjo-Ferrín and Jessica Planelles Villalva and Jose Miguel Torrejón and Jose Joaquín Rodes-Roca},
  title = {xraybinaryorbit: A Python Package for Analyzing Orbital Modulations in X-ray Binaries},
  journal = {Journal of Open Source Software}
}

@ARTICLE{2025A&A...694A.192S,
       author = {{Sanjurjo-Ferr{\'\i}n}, G. and {Torrej{\'o}n}, J.~M. and {Postnov}, K. and {Nowak}, M. and {Rodes-Roca}, J.~J. and {Oskinova}, L. and {Planelles-Villalva}, J. and {Schulz}, N.},
        title = "{Cyclical accretion regime change in the slow X-ray pulsar 4U 0114+65 observed with Chandra}",
      journal = {\aap},
         year = 2025,
        month = feb,
       volume = {694},
          eid = {A192},
        pages = {A192},
          doi = {10.1051/0004-6361/202452789},
archivePrefix = {arXiv},
       eprint = {2501.08702},
 primaryClass = {astro-ph.HE},
       adsurl = {https://ui.adsabs.harvard.edu/abs/2025A&A...694A.192S}
}

@ARTICLE{2001A&A...365L..27T,
       author = {{Turner}, M.~J.~L. and {Abbey}, A. and {Arnaud}, M. and {Balasini}, M. and {Barbera}, M. and {Belsole}, E. and {Bennie}, P.~J. and {Bernard}, J.~P. and {Bignami}, G.~F. and {Boer}, M. and {Briel}, U. and {Butler}, I. and {Cara}, C. and {Chabaud}, C. and {Cole}, R. and {Collura}, A. and {Conte}, M. and {Cros}, A. and {Denby}, M. and {Dhez}, P. and {Di Coco}, G. and {Dowson}, J. and {Ferrando}, P. and {Ghizzardi}, S. and {Gianotti}, F. and {Goodall}, C.~V. and {Gretton}, L. and {Griffiths}, R.~G. and {Hainaut}, O. and {Hochedez}, J.~F. and {Holland}, A.~D. and {Jourdain}, E. and {Kendziorra}, E. and {Lagostina}, A. and {Laine}, R. and {La Palombara}, N. and {Lortholary}, M. and {Lumb}, D. and {Marty}, P. and {Molendi}, S. and {Pigot}, C. and {Poindron}, E. and {Pounds}, K.~A. and {Reeves}, J.~N. and {Reppin}, C. and {Rothenflug}, R. and {Salvetat}, P. and {Sauvageot}, J.~L. and {Schmitt}, D. and {Sembay}, S. and {Short}, A.~D.~T. and {Spragg}, J. and {Stephen}, J. and {Str{\"u}der}, L. and {Tiengo}, A. and {Trifoglio}, M. and {Tr{\"u}mper}, J. and {Vercellone}, S. and {Vigroux}, L. and {Villa}, G. and {Ward}, M.~J. and {Whitehead}, S. and {Zonca}, E.},
        title = "{The European Photon Imaging Camera on XMM-Newton: The MOS cameras}",
      journal = {\aap},
         year = 2001,
        month = jan,
       volume = {365},
        pages = {L27-L35},
          doi = {10.1051/0004-6361:20000087},
archivePrefix = {arXiv},
       eprint = {astro-ph/0011498},
 primaryClass = {astro-ph},
       adsurl = {https://ui.adsabs.harvard.edu/abs/2001A&A...365L..27T}
}

@ARTICLE{2005SSRv..120..143B,
       author = {{Barthelmy}, Scott D. and {Barbier}, Louis M. and {Cummings}, Jay R. and {Fenimore}, Ed E. and {Gehrels}, Neil and {Hullinger}, Derek and {Krimm}, Hans A. and {Markwardt}, Craig B. and {Palmer}, David M. and {Parsons}, Ann and {Sato}, Goro and {Suzuki}, Masaya and {Takahashi}, Tadayuki and {Tashiro}, Makota and {Tueller}, Jack},
        title = "{The Burst Alert Telescope (BAT) on the SWIFT Midex Mission}",
      journal = {\ssr},
         year = 2005,
        month = oct,
       volume = {120},
       number = {3-4},
        pages = {143-164},
          doi = {10.1007/s11214-005-5096-3},
archivePrefix = {arXiv},
       eprint = {astro-ph/0507410},
 primaryClass = {astro-ph},
       adsurl = {https://ui.adsabs.harvard.edu/abs/2005SSRv..120..143B}
}

@ARTICLE{2021Atoms...9...12M,
       author = {{Mendoza}, Claudio and {Bautista}, Manuel A. and {Deprince}, J{\'e}r{\^o}me and {Garc{\'\i}a}, Javier A. and {Gatuzz}, Efra{\'\i}n and {Gorczyca}, Thomas W. and {Kallman}, Timothy R. and {Palmeri}, Patrick and {Quinet}, Pascal and {Witthoeft}, Michael C.},
        title = "{The XSTAR Atomic Database}",
      journal = {Atoms},
         year = 2021,
        month = feb,
       volume = {9},
       number = {1},
          eid = {12},
        pages = {12},
          doi = {10.3390/atoms9010012},
archivePrefix = {arXiv},
       eprint = {2012.02041},
 primaryClass = {astro-ph.IM},
       adsurl = {https://ui.adsabs.harvard.edu/abs/2021Atoms...9...12M}
}

@software{dipspeaks_zenodo,
  author       = {Sanjurjo-Ferr{\'\i}n, G. and others},
  title        = {dipspeaks: detection of dips and spikes in X-ray light curves},
  year         = {2025},
  version      = {x.y.z},
  publisher    = {Zenodo},
  doi          = {10.5281/zenodo.17691607},
  url          = {https://doi.org/10.5281/zenodo.17691607}
}

@ARTICLE{2020A&A...643A...9E,
       author = {{El Mellah}, I. and {Grinberg}, V. and {Sundqvist}, J.~O. and {Driessen}, F.~A. and {Leutenegger}, M.~A.},
        title = "{Radiography in high mass X-ray binaries. Micro-structure of the stellar wind through variability of the column density}",
      journal = {\aap},
         year = 2020,
        month = nov,
       volume = {643},
          eid = {A9},
        pages = {A9},
          doi = {10.1051/0004-6361/202038791},
archivePrefix = {arXiv},
       eprint = {2006.16216},
 primaryClass = {astro-ph.HE},
       adsurl = {https://ui.adsabs.harvard.edu/abs/2020A&A...643A...9E}
}

@article{MartinezNunez2017,
  author  = {Mart{\'i}nez-N{\'u}{\~n}ez, Silvia and Kretschmar, Peter and Bozzo, Enrico and Oskinova, Lidia M. and Puls, Joachim and Sidoli, Lara and Sundqvist, Jon Olof and Blay, Pere and Falanga, Maurizio and F{\"u}rst, Felix and G{\'i}menez-Garc{\'i}a, Angel and Kreykenbohm, Ingo and K{\"u}hnel, Matthias and Sander, Andreas and Torrej{\'o}n, Jos{\'e} Miguel and Wilms, J{\"o}rn},
  title   = {Towards a Unified View of Inhomogeneous Stellar Winds in Isolated Supergiant Stars and Supergiant High Mass X-Ray Binaries},
  journal = {Space Science Reviews},
  year    = {2017},
  volume  = {212},
  number  = {1},
  pages   = {59--150},
  doi     = {10.1007/s11214-017-0340-1},
  url     = {https://doi.org/10.1007/s11214-017-0340-1}
}

@article{Shakura:2018fQ,
  author = "Shakura, N. I.  and  Postnov, Konstantin",
  title = "{Wind Accretion - Observations Vs Theory}",
  doi = "10.22323/1.288.0040",
  journal = "PoS",
  year = 2018,
  volume = "APCS2016",
  pages = "040"
}

@ARTICLE{2017A&A...606L..10B,
       author = {{Bozzo}, E. and {Oskinova}, L. and {Lobel}, A. and {Hamann}, W.-R.},
        title = "{The super-orbital modulation of supergiant high-mass X-ray binaries}",
      journal = {\aap},
         year = 2017,
        month = oct,
       volume = {606},
          eid = {L10},
        pages = {L10},
          doi = {10.1051/0004-6361/201731930},
archivePrefix = {arXiv},
       eprint = {1710.01877},
 primaryClass = {astro-ph.HE},
       adsurl = {https://ui.adsabs.harvard.edu/abs/2017A&A...606L..10B}
}
\newpage

\onecolumn

\begin{appendix}
\FloatBarrier
\section{Average spectra}

\begin{table*}
\caption{Best-fit parameters for the average spectra.}
\label{tab:average_spectra}
\centering
\begin{adjustbox}{max width=\textwidth}
\begin{tabular}{cccccccccc}
\hline\hline
Obs. & MJD & $\chi^2_\nu$ & dof & Orbital phase
& $N_{\rm H}$
& BMC norm
& $kT_{\rm BMC}$
& $\alpha_{\rm BMC}$
& BB norm \\
&&&&
& $(10^{22}\,{\rm cm}^{-2})$
& $(10^{-3}\,L_{39}/D_{10}^2)$
& $(\rm keV)$
& 
& $(10^{-4}\,(R_{\rm km}/D_{10})^2)$ \\
\hline
XMM-1
& 57255.8
& 1.51
& 148
& 0.73--0.75
& $1.30\pm0.04$
& $20.7^{+1.8}_{-2.4}$
& $2.01^{+0.04}_{-0.05}$
& $0.015^{+0.011}_{-0.006}$
& $3.3\pm1.0$ \\ 

\chandra\ Orb 1\tablefootmark{a}   
& 59367.4 
& 1.0  
& --- 
& 0.74--0.76 
& $1.9\pm0.1$            
& $26^{+3}_{-4}$             
& $1.7\pm0.1$            
& $0.018^{+0.004}_{-0.003}$ 
& $3.5^{+1.4}_{-1.3}$ \\ 

\chandra\ Orb 14\tablefootmark{a}  
& 59521.6 
& 1.3  
& --- 
& 0.03--0.41 
& $1.3\pm0.1$            
& $0.25\pm0.01$              
& $1.40^{+0.03}_{-0.05}$ 
& $4.0^{+0.0}_{-0.8}$       
& $2.7\pm0.6$ \\ 

\chandra\ Orb 31\tablefootmark{a}  
& 59719.6 
& 1.2  
& --- 
& 0.11--0.13 
& $2.1\pm0.2$            
& $11.0^{+0.4}_{-0.5}$       
& $1.50\pm0.05$          
& $0.01^{+0.01}_{-0.00}$    
& $2.9\pm1.4$ \\ 

\chandra\ Orb 32\tablefootmark{a}  
& 59729.72 
& 1.5  
& --- 
& 0.99--0.01 
& $23^{+3}_{-2}$         
& $\sim1.6$                  
& $1.8^{+0.2}_{-0.5}$    
& $4^{+0}_{-3}$       
& $<= 5$ \\ 

\swift                              
& 60880.7 
& 1.0 
& 93  
& 0.31--0.33 
& $1.9\pm0.2$            
& $9^{+20}_{-5}$             
& $2.4^{+0.3}_{-0.2}$    
& $0.4^{+2.2}_{-0.5}$       
& $\leq7$ \\ 

XMM-2
& 60890.5
& 1.15
& 157
& 0.07--0.08
& $2.25^{+0.12}_{-0.11}$
& $0.99^{+0.16}_{-0.14}$
& $1.25\pm0.06$
& $0.052^{+0.015}_{-0.013}$
& $1.0\pm0.4$ \\
\hline
\end{tabular}
\end{adjustbox}
\tablefoot{
\tablefoottext{a}{\chandra\ spectral parameters were collected from \cite{2025A&A...694A.192S}, in which the same model was used to fit the spectra.}
}

\end{table*}

\FloatBarrier
\section{Phase-resolved spectra}

\begin{table*}
\caption{Best-fit spectral parameters for the XMM-1 phase-resolved analysis.}
\label{tab:phase_resolved_appendix1}
\centering
\begin{tabular}{ccccccc}
\hline\hline
Bin & $\chi^2_\nu$ & dof & $N_{\rm H}$ & BMC norm & $kT_{\rm BMC}$ & BB norm \\
      &              &     & ($10^{22}\,\mathrm{cm}^{-2}$)
      & ($10^{-3}\,L_{39}/D_{10}^2$)
      & (keV)
      & ($10^{-3}\,(R_{\rm km}/D_{10})^2$) \\
\hline
1  & 1.34 & 107 & $3.4 \pm 0.3$          & $23 \pm 1$             & $2.2 \pm 0.1$ & $0.3 \pm 0.3$ \\ 
2  & 1.12 & 113 & $3.1 \pm 0.4$          & $9 \pm 1$              & $2.1 \pm 0.1$ & $ < 0.4$ \\ 
3  & 1.51 & 128 & $5.4 \pm 0.4$          & $25 \pm 1$             & $2.1 \pm 0.1$ & $ < 0.4$ \\ 
4  & 1.32 & 126 & $7.8 \pm 0.5$          & $35 \pm 2$             & $2.1 \pm 0.1$ & $ < 0.3$ \\ 
5  & 0.97 & 110 & $8.6 \pm 0.7$          & $26 \pm 2$             & $2.3 \pm 0.1$ & $0.3 \pm 0.2$ \\ 
6  & 1.24 & 106 & $2.9^{+0.4}_{-0.3}$    & $9\pm 1$               & $2.2^{+0.2}_{-0.1}$ & $0.3 \pm 0.2$ \\ 
7  & 1.25 & 123 & $3.0 \pm 0.2$          & $21 \pm 1$             & $2.0 \pm 0.1$ & $< 0.6$ \\ 
8  & 1.55 & 136 & $2.9 \pm 0.2$          & $32 \pm 1$             & $1.9 \pm 0.1$ & $0.6 \pm 0.4$ \\ 
9  & 1.29 & 136 & $1.8 \pm 0.1$          & $30 \pm 1$             & $1.9 \pm 0.1$ & $0.5 \pm 0.4$ \\ 
10 & 1.10 & 125 & $1.5 \pm 0.2$          & $8.9 \pm 0.5$          & $1.8 \pm 0.1$ & $0.5 \pm 0.3$ \\ 
11 & 1.06 & 112 & $1.22 \pm 0.08$        & $23 \pm 1$             & $1.8 \pm 0.1$ & $< 0.8$ \\ 
12 & 1.59 & 99  & $0.89 \pm 0.05$        & $34 \pm 1$             & $1.7 \pm 0.1$ & $<0.25$ \\ 
13 & 1.16 & 125 & $0.88 \pm 0.05$        & $30 \pm 1$             & $1.8 \pm 0.1$ & $<0.27$ \\ 
14 & 1.49 & 133 & $0.9\pm 0.1$           & $5.3 \pm 0.3$          & $1.6 \pm 0.1$ & $<0.27$ \\ 
15 & 1.29 & 106 & $0.9 \pm 0.1$          & $16.7 \pm 0.6$         & $1.7 \pm 0.1$ & $<0.47$ \\ 
16 & 1.01 & 84  & $0.96 \pm 0.05$        & $37 \pm 1$             & $1.7 \pm 0.1$ & $<0.24$ \\ 
17 & 1.28 & 94  & $1.2 \pm 0.1$          & $5.6 \pm 0.3$          & $1.5 \pm 0.1$ & $<0.39$ \\ 
18 & 1.02 & 91  & $1.4^{+0.3}_{-0.2}$    & $2.4 \pm 0.2$          & $1.7 \pm 0.1$ & $0.40 \pm 0.21$ \\ 
19 & 1.42 & 110 & $1.06 \pm 0.11$        & $8.7 \pm 0.4$          & $1.7 \pm 0.1$ & $0.4 \pm 0.3$ \\ 
20 & 1.03 & 102 & $1.08 \pm 0.08$        & $15.6 \pm 0.6$         & $1.7 \pm 0.1$ & $0.46 \pm 0.41$ \\ 
21 & 1.13 & 77  & $1.2\pm 0.2$           & $15\pm 1$              & $1.6 \pm 0.1$ & $2.6 \pm 1.2$ \\
\hline
\end{tabular}
\tablefoot{Columns report the reduced $\chi^2$, the number of degrees of freedom (dof), the equivalent hydrogen column density $N_{\rm H}$, the \texttt{bmc} normalization, the \texttt{bmc} temperature $kT_{\rm BMC}$, and the blackbody normalization.}
\end{table*}

\begin{table*}
\caption{Best-fit spectral parameters for the XMM-2 phase-resolved analysis.}
\label{tab:phase_resolved_appendix2}
\centering
\begin{tabular}{ccccccc}
\hline\hline
Bin & $\chi^2_\nu$ & dof & $N_{\rm H}$ & BMC norm & $kT_{\rm BMC}$ & BB norm \\
      &              &     & ($10^{22}\,\mathrm{cm}^{-2}$)
      & ($10^{-3}\,L_{39}/D_{10}^2$)
      & (keV)
      & ($10^{-3}\,(R_{\rm km}/D_{10})^2$) \\
\hline
1 & 1.02 & 66 & $1.9^{+0.8}_{-0.6}$ & $0.33^{+0.05}_{-0.04}$ & $1.3 \pm 0.2$ & $0.4 \pm 0.2$ \\ 
2 & 1.39 & 78 & $2.6^{+0.5}_{-0.4}$ & $1.1 \pm 0.1$           & $1.3 \pm 0.1$ & $0.4 \pm 0.2$ \\ 
3 & 0.91 & 83 & $2.3 \pm 0.3$       & $1.8 \pm 0.1$           & $1.3 \pm 0.1$ & $0.4 \pm 0.2$ \\ 
4 & 1.24 & 87 & $2.4^{+0.5}_{-0.4}$ & $0.9 \pm 0.1$           & $1.2 \pm 0.1$ & $0.5 \pm 0.2$ \\ 
5 & 0.92 & 67 & $1.9^{+0.9}_{-0.6}$ & $0.37^{+0.06}_{-0.05}$  & $1.2 \pm 0.2$ & $0.4 \pm 0.2$ \\ 
6 & 1.02 & 78 & $2.0^{+0.4}_{-0.3}$ & $1.1 \pm 0.1$           & $1.3 \pm 0.1$ & $0.3 \pm 0.2$ \\ 
7 & 1.10 & 24 & $2.0 \pm 0.3$       & $2.0 \pm 0.1$           & $1.3 \pm 0.1$ & $< 0.4$ \\ 
8 & 1.10 & 21 & $2.2 \pm 0.4$       & $1.1 \pm 0.1$           & $1.3 \pm 0.1$ & $0.3 \pm 0.2$ \\ 
9 & 1.49 & 12 & $1.9^{+1.0}_{-0.7}$ & $0.4 \pm 0.1 $          & $1.2 \pm 0.2$ & $0.6 \pm 0.3$ \\

\hline
\end{tabular}
\tablefoot{Columns report the reduced $\chi^2$, the number of degrees of freedom (dof), the equivalent hydrogen column density $N_{\rm H}$, the \texttt{bmc} normalization, the \texttt{bmc} temperature $kT_{\rm BMC}$, and the blackbody normalization.}
\end{table*}

\FloatBarrier

\section{NS-spin folded spectral parameters}
\begin{table*}
\caption{Best-fit spectral parameters for the XMM-1 NS-spin folded spectral analysis.}
\label{tab:ns_spin_folded_xmm1}
\centering
\begin{tabular}{ccccccc}
\hline\hline
Phase & $\chi^2_\nu$ & dof & $N_{\rm H}$ & BMC norm & $kT_{\rm BMC}$ & BB norm \\
      &              &     & ($10^{22}\,\mathrm{cm}^{-2}$)
      & ($10^{-3}\,L_{39}/D_{10}^2$)
      & (keV)
      & ($10^{-3}\,(R_{\rm km}/D_{10})^2$) \\
\hline
0.00 & 1.14 & 136 & $1.30 \pm 0.07$ & $9.9 \pm 0.3$  & $2.01 \pm 0.04$ & $0.09 \pm 0.09$ \\ 
0.05 & 1.24 & 143 & $1.35 \pm 0.06$ & $15.4 \pm 0.4$ & $2.07 \pm 0.03$ & $0.17 \pm 0.12$ \\ 
0.10 & 1.24 & 145 & $1.35 \pm 0.05$ & $21.4 \pm 0.4$ & $2.05 \pm 0.03$ & $0.24 \pm 0.14$ \\ 
0.15 & 1.28 & 148 & $1.27 \pm 0.04$ & $27.4 \pm 0.5$ & $2.06 \pm 0.02$ & $0.26 \pm 0.15$ \\ 
0.20 & 1.48 & 149 & $1.27 \pm 0.04$ & $34.0 \pm 0.5$ & $2.07 \pm 0.02$ & $0.26 \pm 0.18$ \\ 
0.25 & 1.13 & 150 & $1.24 \pm 0.03$ & $38.0 \pm 0.6$ & $2.06 \pm 0.02$ & $0.26 \pm 0.19$ \\ 
0.30 & 1.38 & 150 & $1.18 \pm 0.03$ & $39.3 \pm 0.6$ & $2.06 \pm 0.02$ & $0.24 \pm 0.19$ \\ 
0.35 & 1.33 & 149 & $1.13 \pm 0.03$ & $39.1 \pm 0.6$ & $2.06 \pm 0.02$ & $0.17^{+0.20}_{-0.17}$ \\ 
0.40 & 1.20 & 149 & $1.11 \pm 0.03$ & $38.2 \pm 0.5$ & $2.02 \pm 0.02$ & $0.10^{+0.20}_{-0.10}$ \\ 
0.45 & 1.65 & 150 & $1.02 \pm 0.03$ & $35.5 \pm 0.5$ & $1.98 \pm 0.02$ & $0.08^{+0.19}_{-0.08}$ \\ 
0.50 & 1.58 & 150 & $1.17 \pm 0.03$ & $35.5 \pm 0.5$ & $2.05 \pm 0.02$ & $0.37 \pm 0.19$ \\ 
0.55 & 1.01 & 150 & $1.20 \pm 0.04$ & $33.5 \pm 0.6$ & $2.13 \pm 0.02$ & $0.28 \pm 0.16$ \\ 
0.60 & 0.99 & 148 & $1.29 \pm 0.04$ & $29.4 \pm 0.5$ & $2.14 \pm 0.03$ & $0.33 \pm 0.16$ \\ 
0.65 & 0.99 & 147 & $1.37 \pm 0.05$ & $27.2 \pm 0.6$ & $2.25 \pm 0.03$ & $0.24 \pm 0.14$ \\ 
0.70 & 1.01 & 143 & $1.47 \pm 0.06$ & $23.7 \pm 0.6$ & $2.27 \pm 0.04$ & $0.24 \pm 0.13$ \\ 
0.75 & 1.07 & 143 & $1.55 \pm 0.07$ & $17.1 \pm 0.4$ & $2.18 \pm 0.04$ & $0.23 \pm 0.11$ \\ 
0.80 & 1.19 & 139 & $1.45 \pm 0.07$ & $11.8 \pm 0.3$ & $2.08 \pm 0.04$ & $0.11 \pm 0.09$ \\ 
0.85 & 1.11 & 136 & $1.40 \pm 0.08$ & $9.7 \pm 0.3$  & $2.01 \pm 0.04$ & $0.14 \pm 0.10$ \\ 
0.90 & 1.33 & 135 & $1.36 \pm 0.08$ & $7.5 \pm 0.2$  & $1.97 \pm 0.04$ & $0.10 \pm 0.08$ \\ 
0.95 & 1.34 & 133 & $1.39 \pm 0.08$ & $6.8 \pm 0.2$  & $1.91 \pm 0.04$ & $0.09 \pm 0.09$ \\
\hline
\end{tabular}
\tablefoot{Columns report the reduced $\chi^2$, the number of degrees of freedom (dof), the equivalent hydrogen column density $N_{\rm H}$, the \texttt{bmc} normalization, the \texttt{bmc} temperature $kT_{\rm BMC}$, and the blackbody normalization.}
\end{table*}

\begin{table*}
\caption{Best-fit spectral parameters for the XMM-2 NS-spin folded spectral analysis.}
\label{tab:ns_spin_folded_xmm2}
\centering
\begin{tabular}{ccccccc}
\hline\hline
Phase & $\chi^2_\nu$ & dof & $N_{\rm H}$ & BMC norm & $kT_{\rm BMC}$ & BB norm \\
      &              &     & ($10^{22}\,\mathrm{cm}^{-2}$)
      & ($10^{-3}\,L_{39}/D_{10}^2$)
      & (keV)
      & ($10^{-3}\,(R_{\rm km}/D_{10})^2$) \\
\hline
0.00 & 1.08 & 74  & $1.7^{+0.4}_{-0.3}$ & $0.52 \pm 0.04$ & $1.33 \pm 0.11$ & $0.32 \pm 0.12$ \\ 
0.05 & 1.03 & 86  & $2.0 \pm 0.3$        & $0.79 \pm 0.05$ & $1.34 \pm 0.09$ & $0.36 \pm 0.13$ \\ 
0.10 & 1.28 & 91  & $2.2 \pm 0.3$        & $1.03 \pm 0.06$ & $1.31 \pm 0.08$ & $0.22 \pm 0.12$ \\ 
0.15 & 1.17 & 96  & $2.3 \pm 0.3$        & $1.19 \pm 0.06$ & $1.29 \pm 0.07$ & $0.30 \pm 0.13$ \\ 
0.20 & 1.19 & 98  & $2.3 \pm 0.2$        & $1.40 \pm 0.07$ & $1.29 \pm 0.07$ & $0.21 \pm 0.12$ \\ 
0.25 & 1.28 & 103 & $2.3 \pm 0.2$        & $1.61 \pm 0.08$ & $1.29 \pm 0.07$ & $0.25 \pm 0.13$ \\ 
0.30 & 1.14 & 109 & $2.1 \pm 0.2$        & $1.83 \pm 0.09$ & $1.32 \pm 0.06$ & $0.32 \pm 0.15$ \\ 
0.35 & 0.94 & 93  & $2.1^{+0.3}_{-0.2}$ & $1.88 \pm 0.11$ & $1.26 \pm 0.08$ & $0.36 \pm 0.24$ \\ 
0.40 & 0.88 & 110 & $2.1 \pm 0.2$        & $1.96 \pm 0.09$ & $1.32 \pm 0.06$ & $0.29 \pm 0.16$ \\ 
0.45 & 0.82 & 112 & $2.2 \pm 0.2$        & $1.96 \pm 0.09$ & $1.30 \pm 0.06$ & $0.34 \pm 0.15$ \\ 
0.50 & 0.87 & 111 & $2.2 \pm 0.2$        & $1.83 \pm 0.08$ & $1.29 \pm 0.06$ & $0.37 \pm 0.15$ \\ 
0.55 & 0.81 & 106 & $2.3 \pm 0.2$        & $1.47 \pm 0.07$ & $1.26 \pm 0.07$ & $0.36 \pm 0.14$ \\ 
0.60 & 0.88 & 98  & $2.3 \pm 0.3$        & $1.14 \pm 0.06$ & $1.25 \pm 0.08$ & $0.40 \pm 0.14$ \\ 
0.65 & 1.04 & 86  & $2.2 \pm 0.3$        & $0.92 \pm 0.06$ & $1.28 \pm 0.09$ & $0.35 \pm 0.13$ \\ 
0.70 & 0.99 & 80  & $2.0 \pm 0.4$        & $0.61 \pm 0.05$ & $1.26 \pm 0.11$ & $0.27 \pm 0.13$ \\ 
0.75 & 0.98 & 73  & $1.9 \pm 0.4$        & $0.43 \pm 0.04$ & $1.23 \pm 0.12$ & $0.24 \pm 0.12$ \\ 
0.80 & 0.90 & 83  & $1.9 \pm 0.4$        & $0.40 \pm 0.03$ & $1.21 \pm 0.10$ & $0.25 \pm 0.10$ \\ 
0.85 & 0.90 & 74  & $1.9 \pm 0.4$        & $0.34 \pm 0.03$ & $1.19 \pm 0.11$ & $0.23 \pm 0.10$ \\ 
0.90 & 1.05 & 69  & $1.8 \pm 0.4$        & $0.34 \pm 0.03$ & $1.20 \pm 0.11$ & $0.24 \pm 0.10$ \\ 
0.95 & 0.87 & 73  & $1.8 \pm 0.4$        & $0.41 \pm 0.03$ & $1.28 \pm 0.11$ & $0.29 \pm 0.11$ \\
\hline
\end{tabular}
\tablefoot{Columns report the reduced $\chi^2$, the number of degrees of freedom (dof), the equivalent hydrogen column density $N_{\rm H}$, the \texttt{bmc} normalization, the \texttt{bmc} temperature $kT_{\rm BMC}$, and the blackbody normalization.}
\end{table*}

\end{appendix}

\label{lastpage}
\end{document}